\documentclass{jfm}

\usepackage{graphicx}
\usepackage{epstopdf,epsfig}
\usepackage{newtxtext}
\usepackage{newtxmath}
\usepackage{makecell}
\usepackage{multirow}
\usepackage{natbib}
\usepackage{hyperref}

\newcommand\We{\mbox{\textit{We}}}
\newcommand\Oh{\mbox{\textit{Oh}}}
\newcommand\Reyn{\mbox{\textit{Re}}}
\renewcommand\Re{\mbox{\textit{Re}}}
\newcommand\Ma{\mbox{\textit{Ma}}}
\newcommand\EX{\mbox{\textit{EX}}}
\newcommand\DX{\mbox{\textit{DX}}}
\newcommand\KH{\mbox{\scriptsize{KH}}}
\newcommand\RT{\mbox{\scriptsize{RT}}}
\newcommand{\RomanNumeralCaps}[1]

\title{Droplet breakup in airflow with strong shear effect}

\author{Zhikun Xu,
  Tianyou Wang
 \and Zhizhao Che
\corresp{\email{chezhizhao@tju.edu.cn}}}
\affiliation{State Key Laboratory of Engines, Tianjin University, Tianjin 300072, China
}

\begin{document}
\maketitle

\begin{abstract}
The deformation and breakup of droplets in airflows is important in spray and atomisation processes, but the shear effect in non-uniform airflow is rarely reported. In this study, the deformation and breakup of droplets in a shear flow of air is investigated experimentally using high-speed imaging, digital image processing, and particle image velocimetry. The results show that in airflow with a strong shear effect, the droplet breakup exhibits unique features due to the uplift and stretching produced by the interaction between the deformed droplet and the shear layer. The breakup process can be divided into three stages according to the droplet morphology and the breakup mechanism, namely the sheet breakup, the swing breakup, and the rim breakup stages. Theoretical analysis reveals that the swing breakup is governed by the transverse Rayleigh--Taylor instability. A regime map of the droplet breakup is produced, and the transitions between different regimes are obtained theoretically. The stretching liquid film during the droplet deformation and the fragment size distribution after droplet breakup are analysed quantitatively, and the results show that they are determined by the competition of breakup at different stages affected by the shear. Finally, the effect of the droplet viscosity is investigated, and the viscosity inhibits the droplet breakup in a strong shear airflow.
\end{abstract}

\section{Introduction}\label{sec:sec1}
The disintegration and dispersion of a liquid volume is a phenomenon that exists in many natural and industrial processes, such as raindrop formation in the atmosphere \citep{Villermaux2009Raindrops}, fuel injection in automobiles and aircraft, and pesticide spraying in agriculture. In contrast to primary atomisation of a bulk liquid disintegration, the secondary breakup refers to the process that the droplets formed in primary atomisation break up further into further smaller droplets due to aerodynamic forces in gas streams. The secondary breakup produces numerous small droplets with a large surface-area-to-volume ratio, and has an essential influence on liquid atomisation.

There have been many studies on droplet breakup in uniform airflow, which can be categorised into three main breakup modes \citep{Guildenbecher2009SecondaryAtomization, Hsiang1995BreakupCategory, Zhao2010BagBreakup}---i.e., bag breakup, multimode breakup, shear-stripping breakup---and some other sub-modes such as bag-stamen \citep{Zhao2013BagStamen}, dual-bag \citep{Cao2007DualbagBreakup}, bag-plume \citep{Jain2015SecondaryBreakup}, and plume-shear \citep{Dai2001MultimodeBreakup}. The different breakup modes are determined primarily by the Weber number ($\We_g$) and the Ohnesorge number ($\Oh$), which indicates that the droplet breakup is controlled mainly by the airflow inertia, the droplet viscous force and the surface tension. Moreover, other dimensionless parameters can also affect droplet breakup, including that a larger density ratio ($\rho_d/\rho_g$) tends to cause a lower deformation rate and more intensive fragmentation \citep{Jain2019HighDensity, Yang2016HighDensity}, a lower Reynolds number ($\Reyn$) results in a higher droplet drag coefficient because it is more likely to form a stable vortex ring on the back of the droplet \citep{Jain2019HighDensity, Poon2012ReynoldNumber}, and the Mach number ($\Ma$) changes the pressure distribution on the surface of the droplet and thus changes the shear-stripping breakup morphology \citep{Wang2020MachNumber}. In addition to the droplet breakup modes under different dimensionless parameters, the physical mechanisms governing the droplet breakup are also important for understanding the breakup process. The mechanisms can be summarised briefly as the interface instability (i.e., the competition between Rayleigh-Taylor (RT) instability and Kelvin-Helmholtz (KH) instability on the surface of the droplet \citep{Theofanous2011DropBreakup, Theofanous2008Instability, Theofanous2012ViscousLiquids}), the gas side pressure distribution (i.e., the changes in the position of the flow separation point \citep{Wang2020MachNumber} and the structure of the wake vortex \citep{Flock2012PIV, Meng2018SheetInstability}), and the `internal flow' hypothesis (i.e., the droplet's internal flow governs its breakup \citep{Guildenbecher2009SecondaryAtomization, Jackiw2021InternalFlow}).

In the above studies, the breakup modes and mechanisms are about the breakup of droplets subject to a uniform airflow. However, in many applications, complicated liquid atomisation strategies are employed to obtain desired performance, such as air-assisted impinging jet atomisation, which uses an air jet to promote the atomisation of the liquid sheet formed by impinging liquid jets \citep{Avulapati2013ParticleSzie}, and coaxial swirling atomisation, which uses the interaction of droplets and the swirling vortices to enhance the atomisation process \citep{Rajamanickam2017CoaxialSwirling}. In these complex atomisation strategies, the effects of the airflow on the droplets include not only the uniform aerodynamic force, but also the turbulence effect of the airflow, the effect of the continuously accelerating or decelerating airflow, and the shearing effect of the airflow. There have been some studies discussing the effects of turbulence and the continuously accelerating airflow. In a turbulence flow, the droplet breakup is more intensive and thorough, but only under the large turbulence integral scale (the scale close to the droplet size), or the high turbulence intensity at high average velocity, can the turbulence modify strongly the morphology of the droplet breakup \citep{Jiao2019TurbulentFlows, Zhao2019Turbulence}. By transforming the effect of turbulence on the droplet breakup into the changes in the effective surface tension of the droplet in the spray model, Omidvar \citeyearpar{Omidvar2019Turbulent} developed an improved droplet breakup model based on the effective surface tension, which had good accuracy in predicting the local Sauter mean diameter of spray droplets. In an accelerating airflow, the transient aerodynamic load on the droplets plays a relevant role to anticipate the onset of droplet breakup, and determines the droplet deformation by combining with the residence time on the droplet's trajectory \citep{Garcia2015AcceleratingFlow, Garcia2020AcceleratingFlow, Schmelz2003AcceleratedFlow}. However, there is a lack of understanding of the shear effect of airflow on droplet breakup. In our previous study \citep{Xu2020ShearFlow}, we investigated the shear effect of airflow under the condition of a low Weber number ($12 < \We_g < 20$), and identified a butterfly breakup mode formed by uneven mass distribution during the droplet deformation due to airflow shear. The butterfly breakup mode is governed by the RT instability mechanism, which can be considered as a variation of the bag breakup mode under the influence of shear. In this study, we extend our study to much higher Weber numbers ($\We_g > 100$) accompanied by a strong shear effect, in which the breakup of droplets exhibits more complex morphologies and mechanisms.

It should be pointed out that there have been many studies on the shear-stripping breakup in the literature \citep{Hsiang1995BreakupCategory, Jalaal2014DropletInstabilities, Theofanous2011DropBreakup}. The shearing effect in those studies is the effect of the shear layer formed on the droplet surface by airflow across the droplet. The mainstream airflow is uniform, and the deformation of the droplet is symmetrical relative to the centre of the windward. Different from the shear-stripping breakup in uniform airflow, the airflow in this study has a strong shear effect before it acts on the droplet, and the influence of the shear flow on the droplet deformation is asymmetric relative to the centre of the windward of the droplet, which further leads to completely different results of droplet deformation morphology and fragment size distribution.

The rest of the paper is organised as follows. The experimental details are described in Section \ref{sec:sec2}. The results are analysed in Section \ref{sec:sec3}, including the characterisation of the shear flow, the droplet breakup process, the analysis of swing breakup, the regime map of droplet breakup, the stretching of the liquid film, the size distribution of the fragments, and the effect of the Ohnesorge number. Conclusions are finally drawn in Section \ref{sec:sec4}.

\section{Experimental set-up}\label{sec:sec2}
\begin{figure}
  \centerline{\includegraphics[width=0.8\columnwidth]{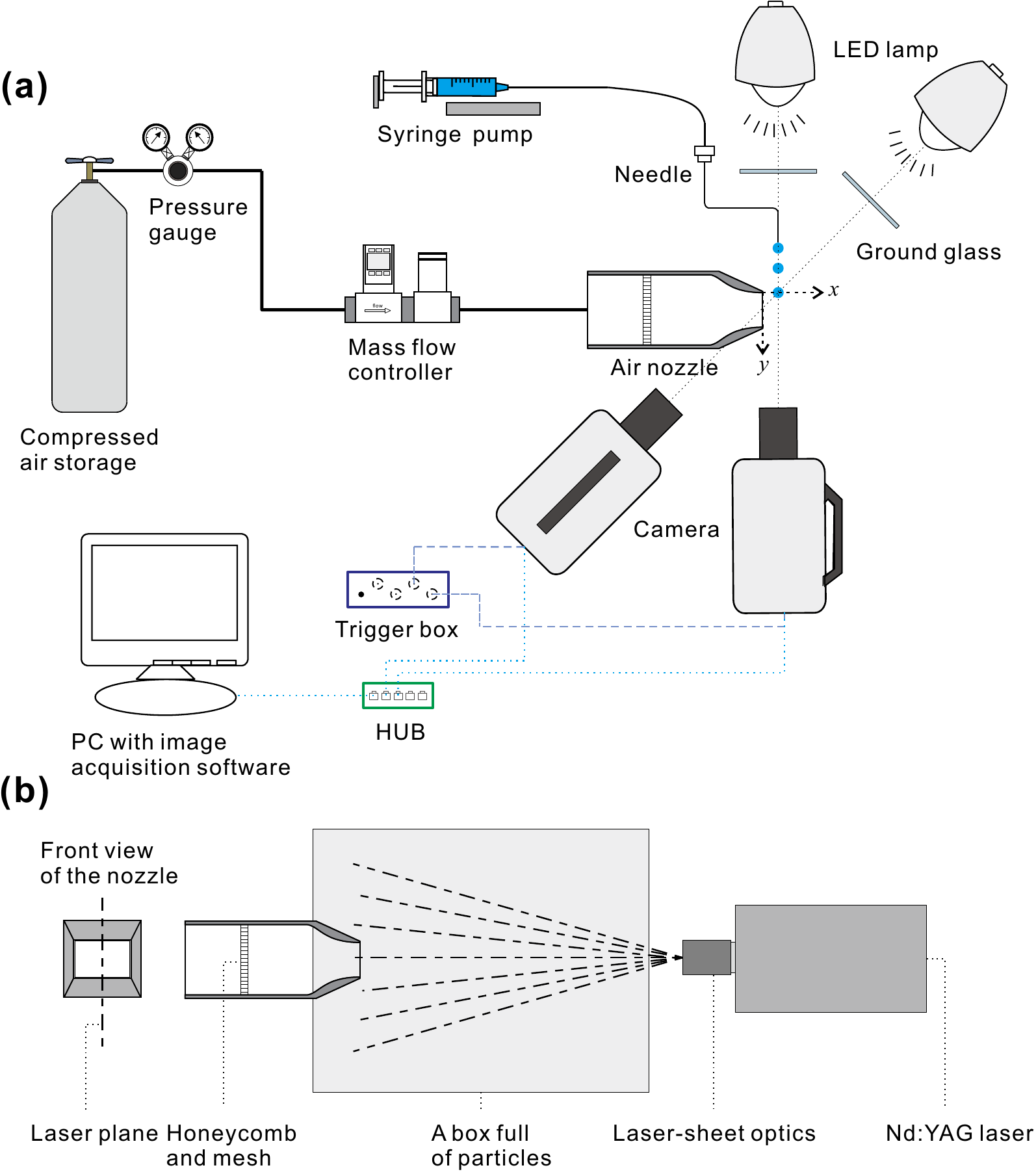}}
  \caption{Schematic diagrams of the experimental set-up. (a) Experimental set-up for high-speed imaging. (b) Experimental set-up for PIV measurements.}
\label{fig:fig01}
\end{figure}

Previous experiments of droplet breakup often used a nozzle jet \citep{Flock2012PIV, Guildenbecher2009SecondaryAtomization, Zhao2011BagBreakup} where the droplet passed through the shear layer of the jet as soon as possible to avoid the effect of the shear layer. In contrast, in this study, we use the shear layer of the air jet to investigate the influence of a strong shear on droplet breakup by adjusting the speed of the droplet passing through the shear layer. The experimental set-up is shown schematically in figure \ref{fig:fig01}a. The air jet was generated from a rectangular nozzle (the outlet cross-section is 30 mm in width and 20 mm in height) combined with a compressed-air cylinder. The airflow rate was adjusted by a mass flow controller (Alicat MCRQ, maximum flow rate 3000 slpm (standard litre per minute), estimated uncertainty $\pm 0.8\%$ of reading and $\pm 0.2\%$ of full scale). After the flow rate adjustment, the airflow passed through a honeycomb and two-layer meshes to reduce turbulence and was ejected from the nozzle. At 3 mm away from the nozzle exit, a droplet entered the shear layer of the air jet under gravity. The droplet was generated at the tip of a blunt needle connected with a syringe pump (Harvard Apparatus, Pump 11 Elite Pico Plus), and the downward speed of the droplet ($u_d$) was tuned by adjusting the falling height of the droplet. The range of droplet downward speed in this study is 0.2--2.5 m/s. The experiments were performed at room temperature (25 $^\circ$C) and atmospheric pressure with air density $\rho_g = 1.2$ kg/m$^3$ and air viscosity $\mu_g = 0.0183$ mPa$\cdot$s. The test liquids were ethanol and silicone oil, whose physical properties are shown in table \ref{tab:tab1}. For the results presented in Sections \ref{sec:sec32}--\ref{sec:sec36}, the test liquid is ethanol, and for Section \ref{sec:sec37}, the test liquids are silicone oils of different viscosities.

To characterise the process of droplet deformation and breakup, several dimensionless parameters are used and varied in the experiments. The gas Weber number (${{\We}_{g}}={{{\rho }_{g}}u_{g}^{2}{{d}_{0}}}/{\sigma }$) was varied in the range 100--917, and the Ohnesorge number ($\Oh={{{\mu }_{d}}}/{\sqrt{{{\rho }_{d}}{{d}_{0}}\sigma }}$) was varied in the range 0.0054--0.95. A droplet falling Weber number (${{\We}_{d}}={{{\rho }_{d}}u_{d}^{2}{{d}_{0}}}/{\sigma }$) was also used in the analysis, and it was varied in the range 3.5--885. A new dimensionless parameter $\chi ={{{u}_{d}}\sqrt{{{\rho }_{d}}}}/{( u_g\sqrt{\rho_g} )}$ is defined to quantify the relative effect of the droplet falling inertia, and can reflect the deflection of the droplet in shear flow (which will be discussed in Section \ref{sec:sec33}). The value of $\chi$ was varied in the range of 0.061--1.964.

\begin{table}
\begin{center}
\def~{\hphantom{0}}
\begin{tabular}{p{2cm}<{\centering}p{2cm}<{\centering}p{2cm}<{\centering}p{2cm}<{\centering}p{2cm}<{\centering}}
Test liquid &\makecell{Viscosity \\($\mu_d$, mPa$\cdot$s)} &\makecell{Density \\($\rho_d$, kg/m$^3$)} &\makecell{Surface tension \\($\sigma$, mN/m)} &\makecell{Diameter \\($d_0$, mm)} \\
Ethanol         & 1.103		& 785		& 21.8   & 2.4 \\
Silicone oil	& 10		& 930		& 20.1   & 2.3\\
Silicone oil	& 50		& 960		& 20.8   & 2.3\\
Silicone oil	& 100		& 966		& 20.9   & 2.2\\
Silicone oil	& 200		& 968		& 21     & 2.2\\
\end{tabular}
\caption{Physical properties of test liquids.}
\label{tab:tab1}
\end{center}
\end{table}

The droplet breakup process was captured by high-speed cameras with different shooting settings for different purposes. Since the droplet breakup in the shear airflow was a non-axisymmetric process, we used two synchronised high-speed cameras (Photron Fastcam SA1.1) to take images from the side and the bottom views, including figures \ref{fig:fig04}, \ref{fig:fig15}, \ref{fig:fig17} and \ref{fig:fig18}. The frame rate of the synchronised cameras was 10000 fps, and the spatial resolution was 40--62 $\mu$m/pixel. To capture the rapidly changing surface fluctuations of the droplet and the rapid fragmentation, we used another high-speed camera (Phantom V2511) with frame rate 20000--40000 fps and spatial resolution 28--40 $\mu$m/pixel to shoot, including figures \ref{fig:fig05}--\ref{fig:fig08} and \ref{fig:fig19}. To obtain the fragment size, we used a 60 mm macro lens (Nikon AF 60 mm f/2.8D) with a small aperture (F22) on the high-speed camera, which can obtain a large depth of field with low optical distortion. We also chose high-power (800 W) light-emitting diode (LED) lights diffused by ground glass as the background light source to ensure sufficient brightness.

Particle image velocimetry (PIV) experiments were conducted to obtain the velocity field of the shear flow of the air jet, as shown in figure \ref{fig:fig01}b. To obtain the flow field structures inside and outside the jet, the air jet was injected into a large empty box ($1\times1\times1.5$ m$^3$). During the experiment, both the jet and the empty box were filled with particles. Particles with an average diameter of about 1--3 $\mu$m were produced by a Laskin sprayer with dioctyl sebacate as the solution, which had good fluidity and was stable in the airflow. A dual-head Nd:YAG laser (SOLO120, 532 nm wavelength) was used to illuminate the particles, and a CCD camera (LaVision Imager SX 4M CCD camera) with image resolution 2360$\times$1776 pixels was used for image acquisition. Spatial resolution 21.8 $\mu$m/pixel was achieved in the acquisition of the particle images. Then the particle images were processed in a PIV system (LaVision). The interrogation window size was iterated from 64$\times$64 pixels to 32$\times$32 pixels with 50\% overlap.

\section{Results and discussion}\label{sec:sec3}
\subsection{Characterisation of shear flow }\label{sec:sec31}
\begin{figure}
  \centerline{\includegraphics[width=1\columnwidth]{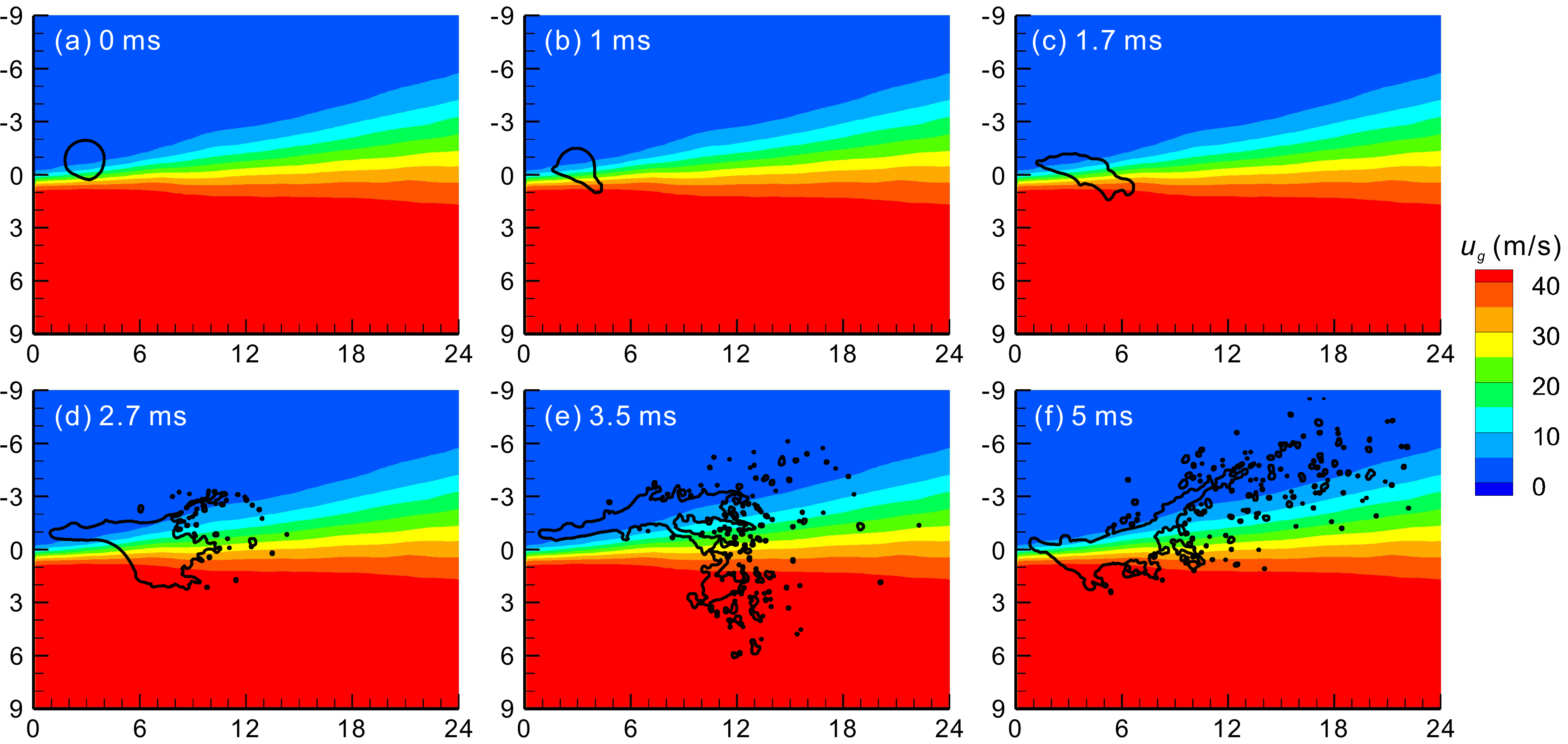}}
  \caption{Average velocity field of the shear flow without droplet breakup and contours of the droplet during the breakup process at jet velocity $u_g=41.7$ m/s. Here, $\We_g = 230$, $\We_d = 19.9$, $\chi = 0.295$, ethanol droplet diameter $d_0 = 2.4$ mm. The values on the axes are the spatial coordinates defined in figure \ref{fig:fig01}.}
\label{fig:fig02}
\end{figure}

A typical velocity field of the jet shear layer obtained by the PIV experiment is shown in figure \ref{fig:fig02}. The velocity field is the average of 400 instantaneous velocity fields, and it has been confirmed in our previous experiment that the average velocity field is stable when the number of instantaneous velocity fields reaches 200 \citep{Xu2020ShearFlow}. From the average velocity field, it can be seen that there is a shear layer with a large velocity gradient between the jet core and the outer region, and this shear layer is used to study the shear effect on droplet breakup.

By controlling the airflow rate, we can change the jet velocity and the shear strength of the shear layer. In our experiments, the range of the jet velocity ($u_g $) is 27.8--83.3 m/s. Correspondingly, the Reynolds number of the jet is $4.4\times10^4$--$1.3\times10^5$, in which the hydraulic diameter (24 mm) of the rectangular nozzle is used as the characteristic length for the Reynolds number calculation. The jet flow is turbulent, and according to the studies by Jiao \emph{et al.} \citeyearpar{Jiao2019TurbulentFlows} and Zhao \emph{et al.} \citeyearpar{Zhao2019Turbulence}, the velocity fluctuation increases the asymmetry of droplet deformation and the randomness of droplet breakup. Therefore, besides qualitative discussions on the influence of shear, statistical data are obtained from repeated experiments to eliminate the impact of turbulence randomness on the data analysis.

\begin{figure}
  \centerline{\includegraphics[width=0.65\columnwidth]{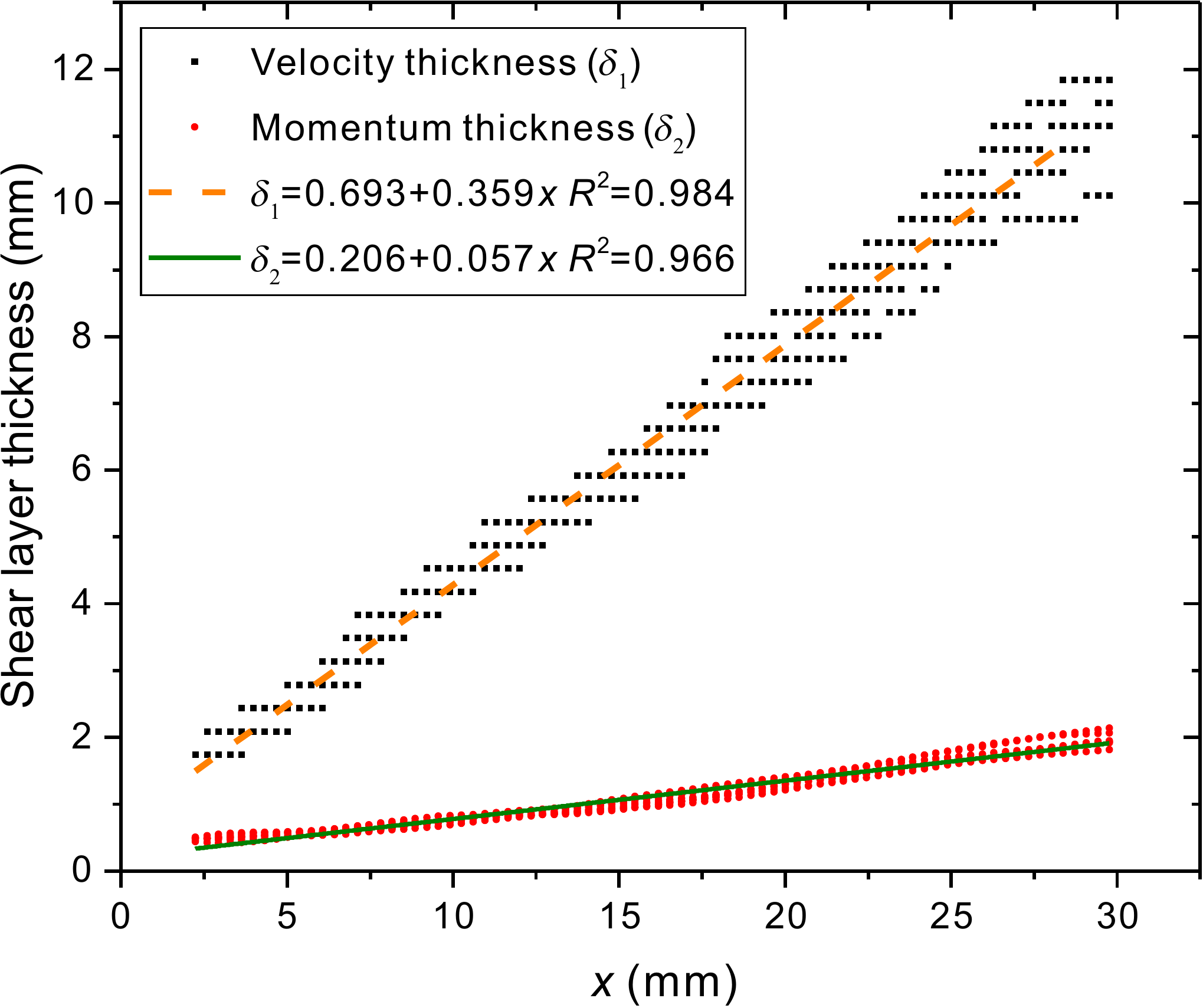}}
  \caption{Velocity thickness ($\delta_1$) and momentum thickness ($\delta_2$) of the shear layer along the flow direction $x$. The range of the jet velocity ($u_g$) is 27.8--83.3 m/s. The lines are linear fitting.}
\label{fig:fig03}
\end{figure}

To further quantify the velocity fields at different jet velocities, we extract the velocity thickness ($\delta_1$) of the shear layer, which is defined as the distance in the $y$-direction from 5\% jet velocity to 95\% jet velocity:
\begin{equation}\label{eq:eq1}
  {{\delta }_{1}}(x)=\int_{{{y}_{1}}\left| u=0.05{{u}_{g}} \right.}^{{{y}_{2}}\left| u=0.95{{u}_{g}} \right.}{1}dy.
\end{equation}
Since the effect of the airflow on the droplet is essentially a momentum transfer process, we also extract the momentum thickness ($\delta_2$) of the shear layer, which is defined as the thickness of a hypothetical inviscid fluid with a uniform velocity that has the same momentum flow rate as the momentum loss rate due to fluid viscosity in the actual shear layer:
\begin{equation}\label{eq:eq2}
  {{\delta }_{2}}(x)=\int_{{{y}_{1}}}^{{{y}_{2}}}{\frac{u(x,y)}{{{u}_{g}}(x)}\left[1-\frac{u(x,y)}{{{u}_{g}}(x)}\right]}dy,
\end{equation}
where the range of integration $y_1< y < y_2$ is some points far from the shear layer. The variations of the velocity thickness and the momentum thickness of the shear layer along the flow direction are shown in figure \ref{fig:fig03}, which shows that both the velocity thickness and the momentum thickness can be considered to increase linearly with $x$, and the thickness of the shear layer at different jet velocities can be regarded as constant.

To understand the relative position of the droplet in the flow field during the droplet breakup process, we extract the droplet contours of a typical breakup process and plot them in the velocity field, as shown in figure \ref{fig:fig02}. It can be seen that in the process of the droplet deformation and breakup, the bottom of the droplet is stretched quickly along the jet direction under the strong shear, while the top of the droplet remains at the original position until it finally breaks. By changing the falling height of the droplet, we can change the downward speed of the droplet entering into the flow field to control the degree of the droplet invasion into the shear flow field, and further change the relative position of the droplet in the flow field.

\subsection{Breakup process}\label{sec:sec32}
\begin{figure}
  \centerline{\includegraphics[width=1\columnwidth]{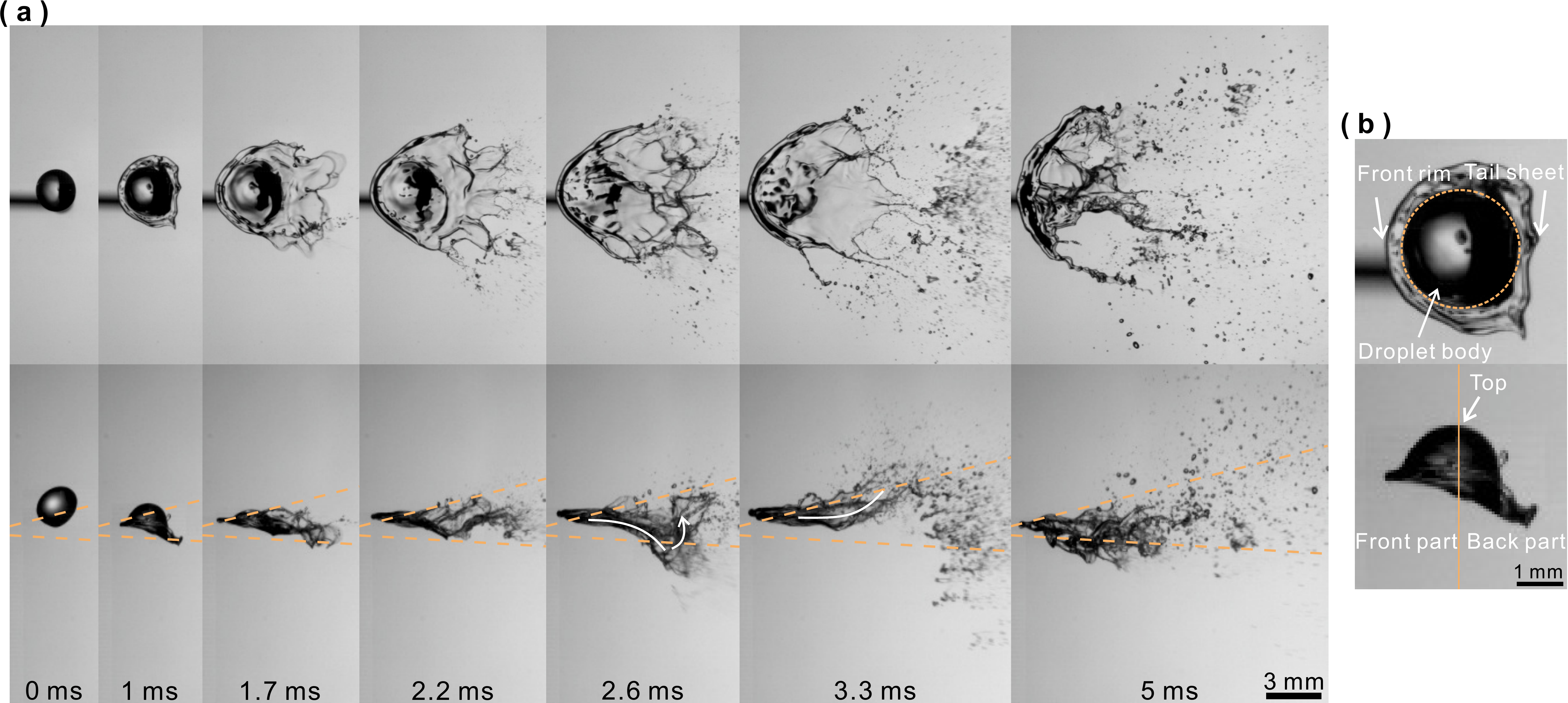}}
  \caption{Image sequences of droplet breakup in shear flow. The direction of the airflow is from left to right: (a) synchronised images from the bottom view (first row) and the side view (second row); (b) enlarged images for the front and the back parts of the droplet at $t = 1$ ms. Here, $\We_g = 230$, $\We_d = 19.9$, $\chi = 0.295$, ethanol droplet diameter $d_0 = 2.4$ mm. The orange lines indicate the velocity boundaries of the shear layer, corresponding to 5--95\% of the jet velocity. The corresponding movie can be found in the supplementary material (movie 1).}
\label{fig:fig04}
\end{figure}

A typical breakup process of a droplet in a strong shear flow is shown in figure \ref{fig:fig04}. We let the droplet enter the jet at a low speed, which makes the droplet break up under the strong shear of the shear layer. As the droplet enters the shear layer, the lower left part of the droplet is first flattened and tilted under the effect of the shear airflow (1 ms in figure \ref{fig:fig04}a). Then the direction of the airflow is deflected on the inclined windward side of the droplet, and a high-pressure area is formed. Similar to the pressure surface (i.e., the lower surface) of a wing, the high-pressure area on the inclined windward side produces an upward lift on the droplet. If we divide the droplet into two parts (the front and back parts as shown in figure \ref{fig:fig04}b), then we can find further that the shear flow has different effects on the front and back parts. At the front part of the droplet, the flattened windward side is almost parallel to the shear layer. Under the combined effect of the downward inertia of the droplet and the upward lifting force of the airflow, a liquid cap appears at the front of the droplet, similar to the impact of a droplet on a wall. At the back part of the droplet, the windward side of the droplet is tilted in the shear layer at a certain angle, and a liquid sheet is stretched out from the tail of the droplet under the strong shear effect.

As the lower part of the droplet is flattened and the upper part of the droplet moves downwards, the main body of the droplet deforms continuously (1.7--2.2 ms in figure \ref{fig:fig04}a). In the uniform flow of a high Weber number, the main body of the droplet is flattened as a consequence of the non-uniform pressure distribution around the droplet surface (high pressure at the front and rear stagnation points, as well as low pressure at the equator due to the acceleration of the gas) \citep{Meng2018SheetInstability}. In contrast, the deformation mechanism of the droplet in shear flow is different, which includes not only the flattening caused by the lifting force of the shear layer and the downward inertia of the droplet, but also the stretching by the shearing force along the flow direction. The combined effects make the main body of the droplet deform into a fan shape (2.2 ms in figure \ref{fig:fig04}a).

During the deformation of the droplet's main body, the liquid cap in the front part of the droplet develops into a stable rim and maintains its position in the flow field. The liquid sheet stretched out from the droplet tail in the early stage breaks continuously on a small scale. When the deformation of the main body is over, the droplet is fully stretched. Since the back part of the droplet has a larger windward area and contains more liquid at this time than the early tail sheet breakup, the deformed droplet will break up intensively from the droplet’s back part (2.6 ms in figure \ref{fig:fig04}a). The intensive breakup at the droplet's back part will cause the droplet to swing upwards remarkably (as highlighted by the arrow at 2.6--3.3 ms in figure \ref{fig:fig04}a). Finally, depending on the strength of the swing, the front rim of the droplet moves upwards out of the flow field driven by the swing, or the rim falls into the high-speed region of airflow by gravity and breaks up because the lifting force decreases with the disintegration of the liquid film. In addition to figure \ref{fig:fig04}, more image sequences at different conditions are provided in Appendix \ref{sec:AppA}.

To analyse further the breakup process of the droplet under the strong shear effect, we divide the breakup process into three stages, according to the interaction between the droplet and airflow: the sheet breakup stage (i.e., the liquid sheet is stretched out from the droplet tail), the swing breakup stage (i.e., the main body of the droplet breaks up), and the rim breakup stage (i.e., the front rim of the droplet collapses or shrinks).

\emph{Sheet breakup stage}: When the droplet just enters the shear flow, the liquid sheet is stretched out from the lower surface of the droplet, which is the main feature of the sheet breakup stage. The stretching of the sheet is similar to the early deformation of the droplet breakup process in a uniform airflow. But in the shear flow, the sheet stretching is asymmetric and occurs mainly at the droplet tail where the shear effect is strong, while it is symmetrical and occurs along the periphery of the droplet in a uniform flow.

Most of the previous studies \citep{Jalaal2014DropletInstabilities, Sharma2021Aerobreakup, Theofanous2011DropBreakup, Theofanous2012ViscousLiquids}, especially those based on shock tube experiments, indicated that the early sheet originated from the KH instability wave on the surface of the droplet. The KH instability wave caused by the shear layer on the droplet surface was further stretched to form the sheet under the airflow action. In a recent study, Jackiw and Ashgriz \citeyearpar{Jackiw2021InternalFlow} attributed the mechanism of the sheet formation to the local normal pressure of the airflow; i.e.\ an initial rim around the droplet formed due to the outward flow of the windward face, then the rim was pushed downstream, and a thin sheet was produced to connect the rim to the main body of the droplet. The difference between the two interpretations depends essentially on whether more emphasis is on the shear effect or the normal pressure of the airflow, which is affected by the experimental conditions. Moreover, both interpretations are based on the analysis of the interplay between the aerodynamic force and the surface tension \citep{Jackiw2021InternalFlow, Villermaux2020FragmentationCohesion, Villermaux1998ShearInstibility}. Therefore, the two interpretations are closely related. From our experimental phenomenon, the early stretching of the sheet is more consistent with the interpretation of Jackiw and Ashgriz \citeyearpar{Jackiw2021InternalFlow}. An initial rim is formed on the periphery of the flattened windward side, then the rim is drafted downstream to form a stretched sheet. The sheet, as it is stretched, becomes thinner and increasingly sensitive to instabilities until it breaks up.

\begin{figure}
  \centerline{\includegraphics[width=0.9\columnwidth]{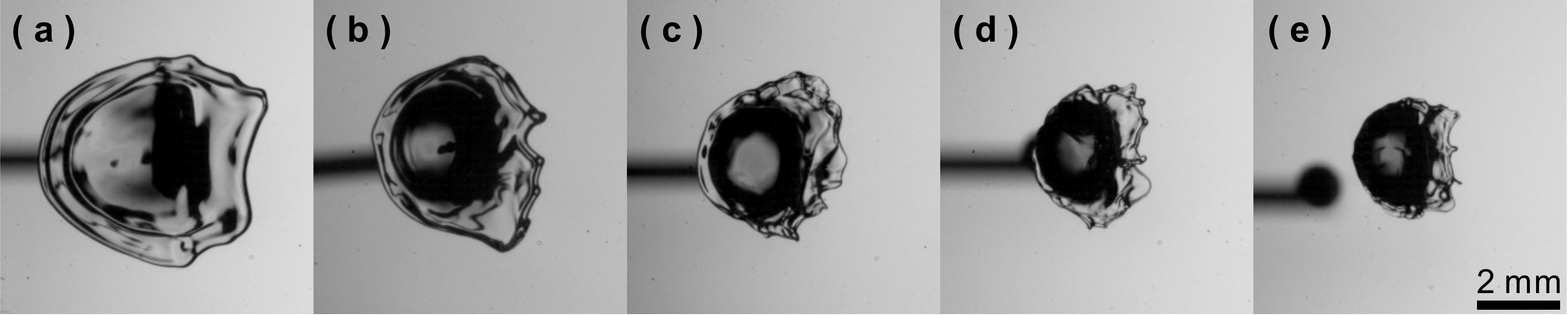}}
  \caption{Bottom-view images of the tail sheet before its first breakup: (a) $\We_g$ = 102, $\chi$ = 0.442; (b) $\We_g$ = 230, $\chi$ = 0.295; (c) $\We_g$ = 408, $\chi$ = 0.221; (d) $\We_g$ = 637, $\chi$ = 0.177; (e) $\We_g$ = 917, $\chi$ = 0.147. Here, $\We_d$ = 19.9, ethanol droplet diameter $d_0 = 2.4$ mm.}
\label{fig:fig05}
\end{figure}

Based on the pressure balance at the initial rim between the Laplace pressure (${2\sigma }/{{{h}_{\text{init}}}}$) and the airflow pressure (${{{\rho }_{g}}u_{g}^{2}}/{2}$), we can estimate the thickness of the rim ${{{h}_{\text{init}}}}/{{{d}_{0}}}\sim {4}/{ {{\We}_{g}}}$. As $\We_g$ increases, the thickness of the initial rim decreases due to the stronger aerodynamic force of the air. Hence the thickness of the stretched sheet is thinner at higher $\We_g$. Figure \ref{fig:fig05} shows the images of the tail sheet before its first breakup under different $\We_g$ with the same $\We_d$. Although different $\We_g$ and $\chi$ lead to different deformation of the droplet body (discussed in detail in Section \ref{sec:sec33}), it can be seen that as $\We_g$ increases, the thinner sheet in the early stage becomes more unstable and is prone to break up quickly under the stronger shear. Thus the spreading area of the sheet becomes smaller as $\We_g$ increases.

\begin{figure}
  \centerline{\includegraphics[width=0.6\columnwidth]{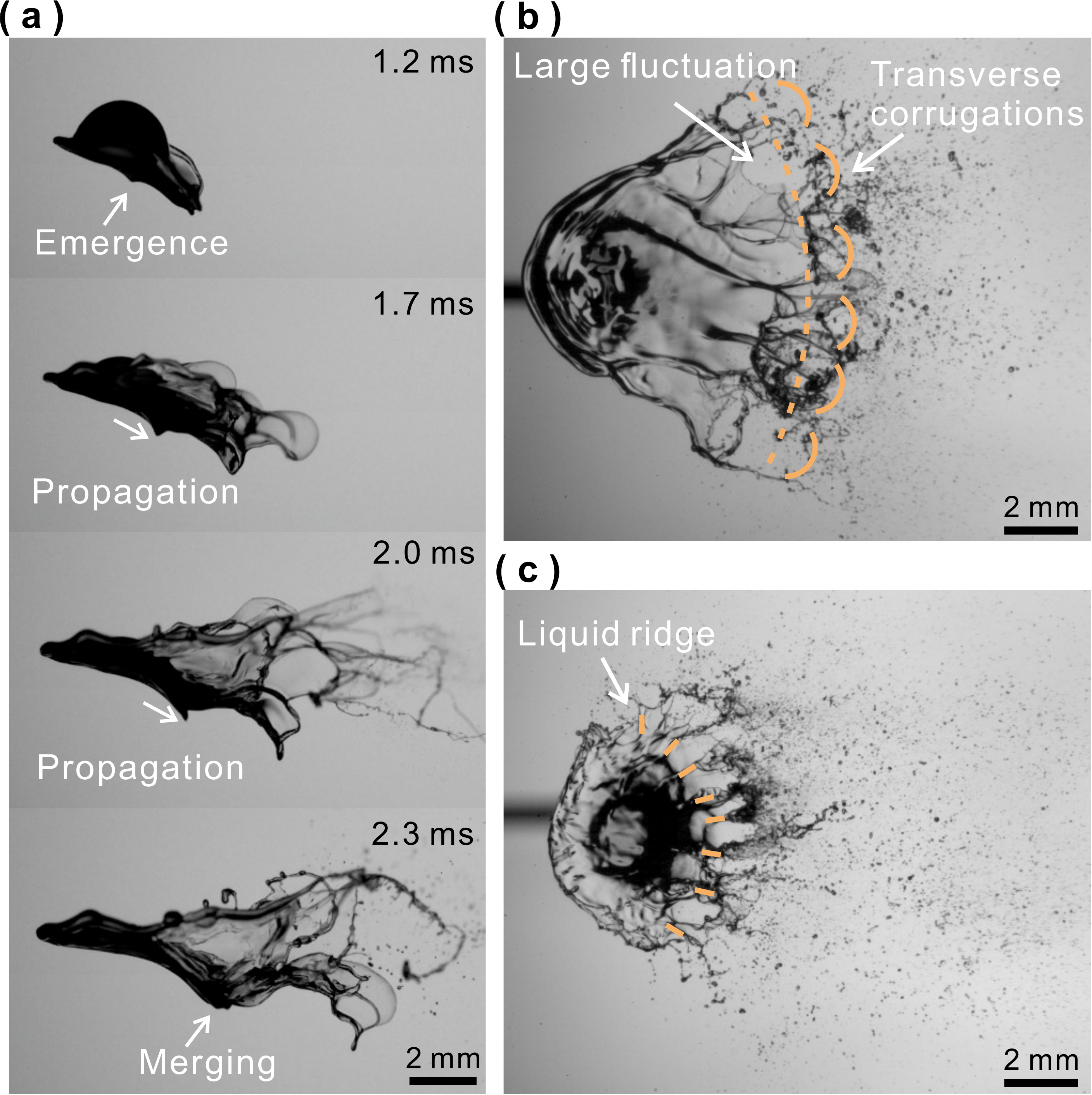}}
  \caption{(a) Emergence, propagation and merging of the surface wave at the droplet windward side in the swing breakup stage in the side-view images: (b) Wave structures in the bottom-view images at $\We_g = 230$, $\We_d = 19.9$, $\chi = 0.295$, ethanol droplet with $d_0 = 2.4$ mm. (c) Wave structures in the bottom-view images at $\We_g = 917$, $\We_d =36.5$, $\chi = 0.199$, ethanol droplet diameter $d_0 = 2.4$ mm (movie 2 in the supplementary material).}
\label{fig:fig06}
\end{figure}

\emph{Swing breakup stage}: The second stage of the breakup features the intensive swing of the droplet body. With the stretching and breakup of the tail sheet, the main body of the droplet is flattened continuously. As the droplet body deforms, the windward surface area of the droplet increases and the thickness of the droplet decreases. Hence the liquid tongue has a larger acceleration. In addition, as the droplet falls and deforms, the windward surface at the tail of the droplet body becomes more perpendicular to the incoming flow. These factors promote the development of the RT instability and the formation of large fluctuation with transverse corrugations on the liquid tongue at the tail of the droplet body, as shown in figure \ref{fig:fig06}b. Even at a much higher Weber number ($\We_g = 917$), the sheet breakup of the tail edge of the droplet in the early stage is more intensive and thorough, but the liquid ridge formed by the lateral RT instability can still be seen on the edge of the droplet body, as shown from the bottom-view image in figure \ref{fig:fig06}c. Further modelling analysis of the RT instability will be given in Section \ref{sec:sec33}. Since this large liquid tongue contains much liquid and the droplet body is thin, the droplet will swing remarkably, driven by the breakup of the liquid tongue.

\begin{figure}
  \centerline{\includegraphics[width=0.8\columnwidth]{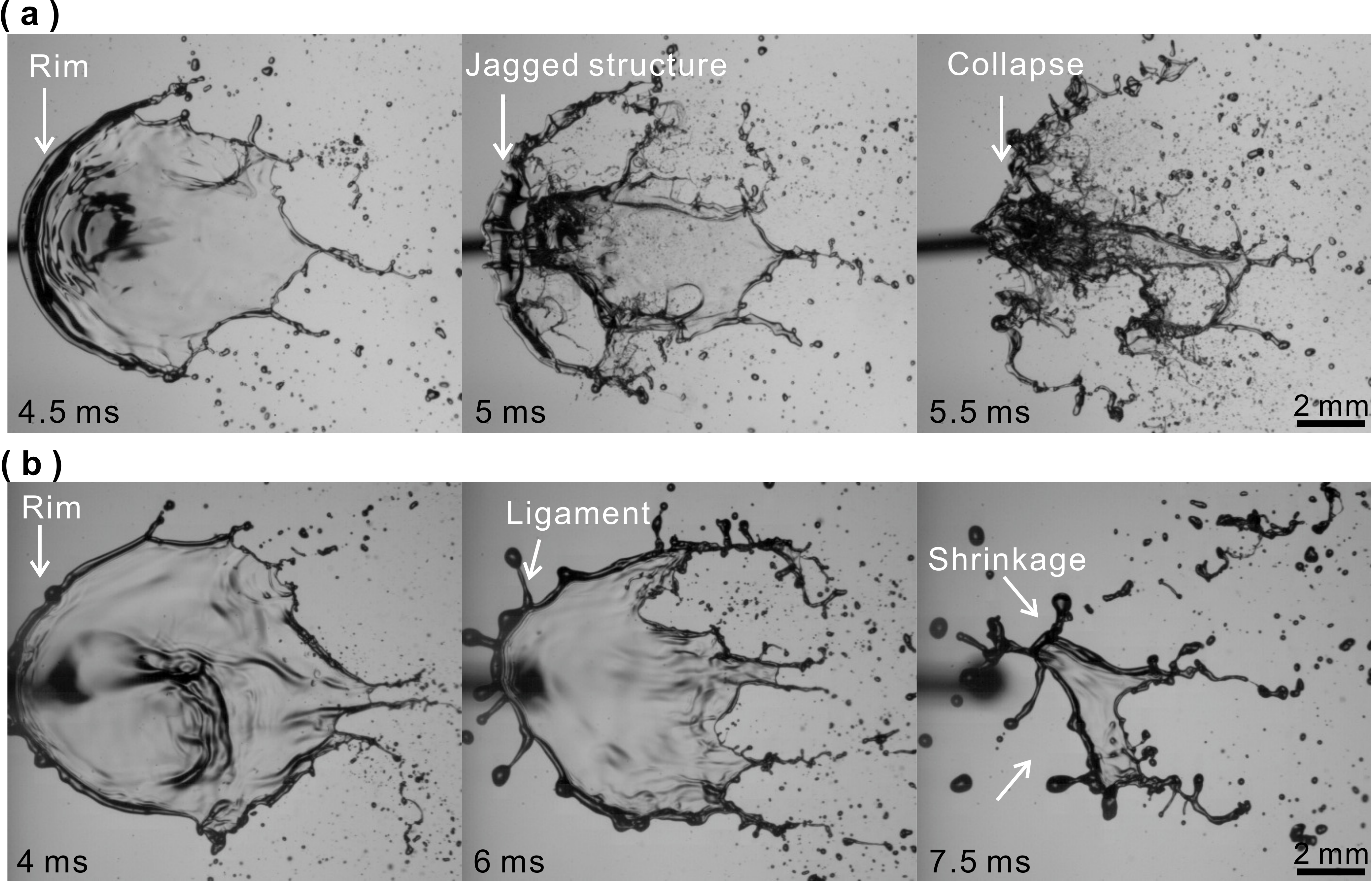}}
  \caption{(a) Collapse of the droplet front rim, $\We_g$ = 230, $\We_d$ = 19.9, $\chi = 0.295$, ethanol droplet diameter $d_0 = 2.4$ mm. (b) Shrinkage of the droplet front rim, $\We_g$ = 230, $\We_d =36.5$, $\chi = 0.399$, ethanol droplet diameter $d_0 = 2.4$ mm. The corresponding movies can be found in the supplementary material (movie 3 for rim collapse, and movie 4 for rim shrinkage). }
\label{fig:fig07}
\end{figure}

\emph{Rim breakup stage}: After the swing breakup stage, the droplet breakup process enters the rim breakup stage, which features mainly the collapse or shrinkage of the front rim of the droplet. In the first two stages, when sheet breakup and swing breakup occur at the back part of the droplet, the front rim of the droplet is still in the region with low air velocity and maintains a stable position due to the uplift effect of the shear layer. But after the swing breakup stage, there are two scenarios for the rim. In the first scenario, the swing breakup stage has consumed only a little liquid and much liquid is left on the rim, and the strength of the swing effect cannot pull the thick rim out of the shear layer. With the reduction of the lifting force of the shear layer caused by the disintegration of the liquid film, the rim enters the high-speed region of the airflow under gravity, and then is penetrated by the RT instability waves. The front of the droplet appears jagged and eventually collapses, as shown in figure \ref{fig:fig07}a. In contrast, in the second scenario, the swing breakup stage has consumed much liquid and a little liquid is left on the rim, and the front rim will be driven out of the shear layer by the strong swing effect. After leaving the air jet, ligaments appear on the rim, and the rim shrinks and forms large fragments, as shown in figure \ref{fig:fig07}b.

It should be noted that although the droplet breakup under the strong shear effect involves the above three subsequential stages, these three stages of breakup compete with each other due to the conservation of mass. If one stage consumes much liquid and its breakup is intensive, then the other two stages will be weak. For example, when the shear effect is very strong, liquid sheets will be peeled off from the tail of the droplet repeatedly. In this situation, the strength of the swinging breakup stage will be reduced.

\subsection{Analysis of swing breakup}\label{sec:sec33}
\begin{figure}
  \centerline{\includegraphics[width=0.55\columnwidth]{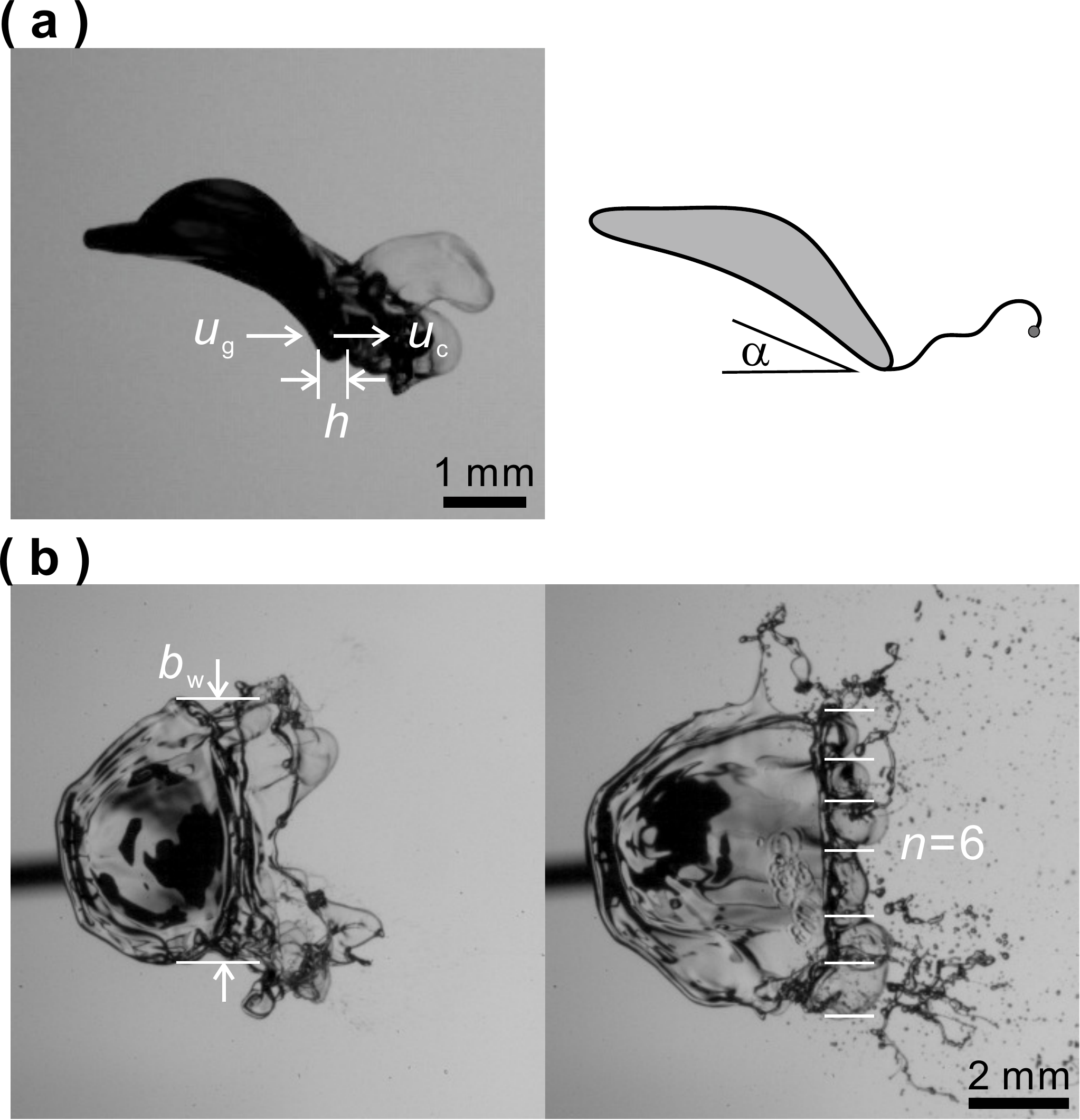}}
  \caption{(a) Acceleration of the liquid tongue. (b) Width of the liquid tongue ($b_w$) before the instability development, and number ($n$) of bags or transverse corrugations after the instability development. Here, $\We_g = 230$, $\We_d = 11.8$, $\chi = 0.227$, ethanol droplet diameter $d_0 = 2.4$ mm.}
\label{fig:fig08}
\end{figure}

The swing breakup process is similar to the flapping breakup of coaxial liquid jets, during which the liquid jet with a coaxial airflow becomes unstable and deviates from its axis to flap \citep{Delon2018Flapping, Aliseda2021Flapping}. However, different from the flapping breakup in coaxial liquid jets, which originates from the shear instability, the swing breakup in this study is induced by the flattening deformation of the droplet body (the detailed comparison is presented in Appendix \ref{sec:AppB}). The droplet body is flattened under the combined effects of the lifting force of the shear layer and the downward inertia of the droplet. Similar to inertia-dominated droplet collisions for low-viscosity liquids \citep{David2004Deformation, Opfer2014FilmThickness, Roisman2009DeformationViscous, Roisman2009Defprmation}, the flattened thickness of the droplet body ($h$) is
\begin{equation}\label{eq:eq3}
  	\frac{h}{{{d}_{0}}}\sim \We_{f}^{-1/2},
\end{equation}
where $\We_f$ is a Weber number for the flattening of the droplet, indicating the ratio of the flattening force to the surface tension. The flattening force consists of the downward inertia of the droplet (${{F}_{d}}\sim {{\rho }_{d}}u_{d}^{2}d_{0}^{2}$) and the lifting force of the shear layer ($F_l$). Since the windward surface of the droplet is tilted in the airflow like a wing, the lifting force can be estimated according to the wing lift theory \citep{Cohen2008FluidMechanics} as
\begin{equation}\label{eq:eq4}
  	{{F}_{l}}\sim {{\rho }_{g}}u_{g}^{2}d_{0}^{2}\sin \alpha ,
\end{equation}
where $\alpha$ is the angle of attack of a wing, which corresponds to the angle of droplet inclination, as shown in figure \ref{fig:fig08}a. The angle of droplet inclination $\alpha$ can be obtained from the comparison of the vertical movement of the droplet (scaled by $u_d$) and the horizontal movement of the droplet (scaled by ${{u}_{g}}\sqrt{{{{\rho }_{g}}}/{{{\rho }_{d}}}}$). For small $\alpha$, we have
\begin{equation}\label{eq:eq5}
  	\sin \alpha \approx \tan \alpha =\frac{{{u}_{d}}}{{{u}_{g}}\sqrt{{{\rho }_{g}}/{{\rho }_{d}}}}=\chi ,
\end{equation}
which indicates that the angle of droplet inclination can be represented by $\chi$. Using (\ref{eq:eq4}) and (\ref{eq:eq5}), we can obtain the flattening Weber number
\begin{equation}\label{eq:eq6}
{{ \We}_{f}}=\frac{\left( {{\rho }_{d}}u_{d}^{2}+{{u}_{g}}{{u}_{d}}\sqrt{{{\rho }_{g}}{{\rho }_{d}}} \right){{d}_{0}}}{\sigma }= {{ \We}_{g}}\left( {{\chi }^{2}}+\chi  \right).
\end{equation}
By comparing (\ref{eq:eq6}) with the original definition of the Weber number, we can obtain the droplet flattening velocity ($u_f$) as
\begin{equation}\label{eq:eq7}
  	{{u}_{f}}=\sqrt{u_{d}^{2}+{{u}_{g}}{{u}_{d}}\sqrt{{{{\rho }_{g}}}/{{{\rho }_{d}}}}}.
\end{equation}
Using this droplet flattening velocity, we can obtain the flattening Reynolds number (${{\Re}_{f}}={{\rho }_{d}}{{d}_{0}}{{u}_{f}}/{{\mu }_{d}}$) and a scaling parameter ($P={{\We}_{f}}\Re_{f}^{-4/5}$), where $P$ is a dimensionless parameter to quantify the relative importance of the inertia used in droplet collision. For the low-viscosity fluid (the test liquid for the results presented in Sections \ref{sec:sec32}--\ref{sec:sec36} is ethanol), ${{\We}_{f}}<430$, ${{\Re}_{f}}>1253$ and $P<0.6$, so the inertia dominates the droplet deformation, and (\ref{eq:eq3}) is valid \citep{David2004Deformation, Roisman2009DeformationViscous, Roisman2009Defprmation}. From (\ref{eq:eq3}) and (\ref{eq:eq6}), we can predict the dimensionless thickness of the droplet body:
\begin{equation}\label{eq:eq8}
  	H=\frac{h}{{{d}_{0}}}={{C}_{1}}\We_{g}^{-1/2}{{\left( {{\chi }^{2}}+\chi  \right)}^{-1/2}},
\end{equation}
where $C_1$ is a constant.

As the RT instability develops, the flattening deformation of the droplet body leads further to the swing breakup. For inviscid fluids, the dispersion equation of the RT instability wave \citep{Chandrasekhar2013Instability} is
\begin{equation}\label{eq:eq9}
  	n_{\text{RT}}^{2}=ak\left[ \frac{{{\rho }_{d}}-{{\rho }_{g}}}{{{\rho }_{d}}+{{\rho }_{g}}}-\frac{{{k}^{2}}\sigma }{a({{\rho }_{d}}+{{\rho }_{g}})} \right],
\end{equation}
where $n_{\RT}$ and $k$ are the growth rate and the wavenumber of the RT instability wave, and $a$ is the acceleration of the liquid. The growth rate ($n_{\RT}$) has a maximum value, and the corresponding most-amplified wavelength is
\begin{equation}\label{eq:eq10}
  	{{\lambda }_{\RT}}=\frac{2\pi }{{{k}_{m}}}=2\pi \sqrt{\frac{3\sigma }{a({{\rho }_{d}}-{{\rho }_{g}})}},
\end{equation}
where $k_m$ is the most-amplified wavenumber. In the early stage of the swing breakup (as shown in figure \ref{fig:fig08}a), the acceleration of the liquid tongue can be considered as
\begin{equation}\label{eq:eq11}
  	a=\frac{{{F}_{D}}}{m}=\frac{{{F}_{D}}}{{{\rho }_{d}}h{{A}_{e}}},
\end{equation}
where $A_e$ is the projected area, $h$ is the thickness of the liquid tongue of the droplet, and $F_D$ is the drag force by the gas on the liquid tongue:
\begin{equation}\label{eq:eq12}
  	{{F}_{D}}=\frac{1}{2}{{C}_{D}}{{\rho }_{g}}{{({{u}_{g}}-{{u}_{c}})}^{2}}{{A}_{e}},
\end{equation}
where $C_D \approx 2$ \citep{White2003FluidMechanics}, and $u_c$ is the backward velocity of the liquid tongue, $u_c \ll u_g$.

\begin{figure}
  \centerline{\includegraphics[width=0.7\columnwidth]{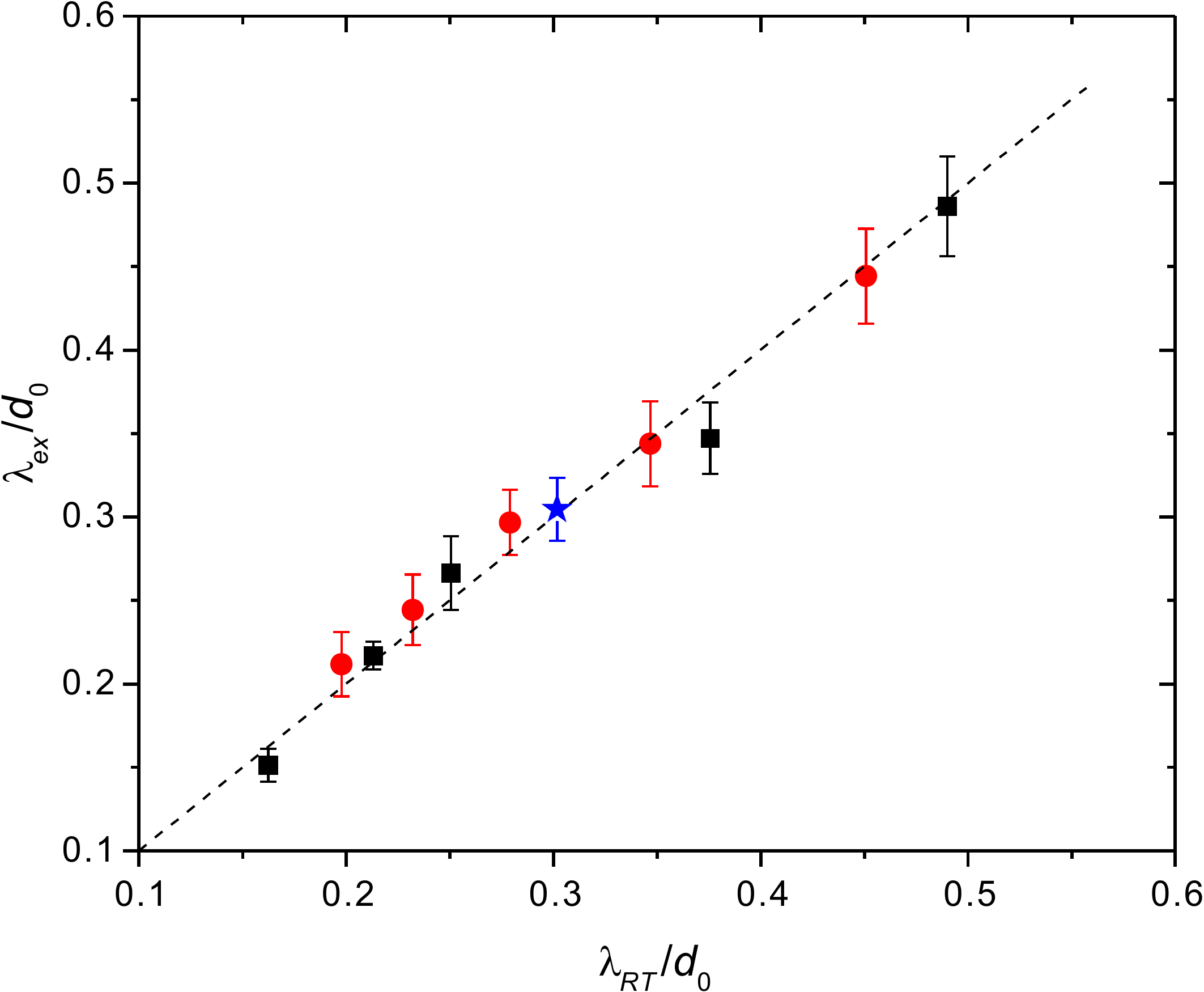}}
  \caption{Comparison between theoretical and experimental instability wavelengths, ${{\lambda }_{\RT}}$ and ${{\lambda }_{ex}}$. The experimental instability wavelength is obtained from experimental images, and the theoretical instability wavelength is obtained from (\ref{eq:eq13}) with a fitted constant $C_1 = 1.41$. The dashed line indicates ${{\lambda }_{RT}}={{\lambda }_{ex}}$. As the wavelength increases, the five black squares correspond to $\We_g$ = 637, 408, 312, 159, 102, respectively, $\chi$ = 0.136, 0.170, 0.195, 0.273, 0.341, respectively, and $\We_d$ = 11.8, ethanol droplet diameter $d_0$ = 2.4 mm. The five red circles correspond to $\We_g$ = 408, 312, 230, 159, 102, respectively, $\chi$ = 0.221, 0.253, 0.295, 0.354, 0.442, respectively, and $\We_d$ = 19.9, ethanol droplet diameter $d_0$ = 2.4 mm. The blue star corresponds to the case of figure \ref{fig:fig08} with $\We_g = 230$, $\chi = 0.227$, $\We_d = 11.8$, ethanol droplet diameter $d_0 = 2.4$ mm.}
\label{fig:fig09}
\end{figure}

By substituting (\ref{eq:eq8}), (\ref{eq:eq11}) and (\ref{eq:eq12}) into (\ref{eq:eq10}), we can deduce the most-amplified wavelength of the RT instability
\begin{equation}\label{eq:eq13}
  	\frac{{{\lambda }_{\RT}}}{{{d}_{0}}}=2\pi \sqrt{3}C_{1}^{1/2}\We_{g}^{-3/4}{{\left( {{\chi }^{2}}+\chi  \right)}^{-1/4}}.
\end{equation}
From (\ref{eq:eq13}), we can obtain the theoretical wavelength of the RT instability wave. For comparison, the experimental wavelength of the RT instability wave can be obtained from the high-speed images as
\begin{equation}\label{eq:eq14}
  	{{\lambda }_{ex}}=\frac{{{b}_{w}}}{n},
\end{equation}
where $b_w$ is the width of the liquid tongue before the development of the instability wave, and $n$ is the number of bags or transverse corrugations after the development of the instability wave (as illustrated in figure \ref{fig:fig08}b). A comparison between the theoretical and experimental wavelengths of the RT instability is made in figure \ref{fig:fig09}, which shows that (\ref{eq:eq13}) can predict the experimental wavelength well.

Although the above instability analysis is based on the swing breakup of droplets in a strong shear airflow, it can also provide an understanding of droplet breakup in uniform airflow due to the similar physical mechanisms involved. In the study of droplet breakup in uniform airflow, following the theory of Villermaux \emph{et al.}\ \citeyearpar{Villermaux2004Transverse} on coaxial jets, Jalaal \emph{et al.}\ \citeyearpar{Jalaal2014DropletInstabilities} believed that the liquid tongue that generates the RT instability was a KH instability wave formed by the development of shear instability (i.e., $h\sim {{\lambda }_{\KH}}$). They obtained the acceleration of the liquid tongue through KH instability analysis, but their theoretical data and simulated data did not match well. From our experimental phenomenon, except for the breakup of the early stretched liquid film, in the further breakup, such as the recurrent breakup proposed by Dorschner \emph{et al.}\ \citeyearpar{Dorschner2020TransverseRTinstability} and the swing breakup in this study, the thickness and the acceleration of the liquid tongue that generates the RT instability is weakly correlated with the KH instability due to the deformation of the droplet body and the merging of the KH instability waves. However, in uniform airflow, it is difficult to obtain directly the thickness of the droplet tongue that generates the RT instability since the stretched liquid film or the droplet mist at the edge of the droplet often blocks the view. In our experiment, the thickness of the liquid tongue is predicted from analogising the flattening deformation of the droplet body in the shear flow to inertia-dominated droplet collisions and referring to the wing lift theory. In addition, we can measure the experimental wavelength through bottom-view images due to the unidirectional stretch effect of the shear layer. Therefore, quantitative results of the RT instability can be obtained, and a good agreement between the experimental data and the theoretical analysis is achieved.

\subsection{Regime map of droplet breakup}\label{sec:sec34}
\begin{figure}
  \centerline{\includegraphics[width=0.65\columnwidth]{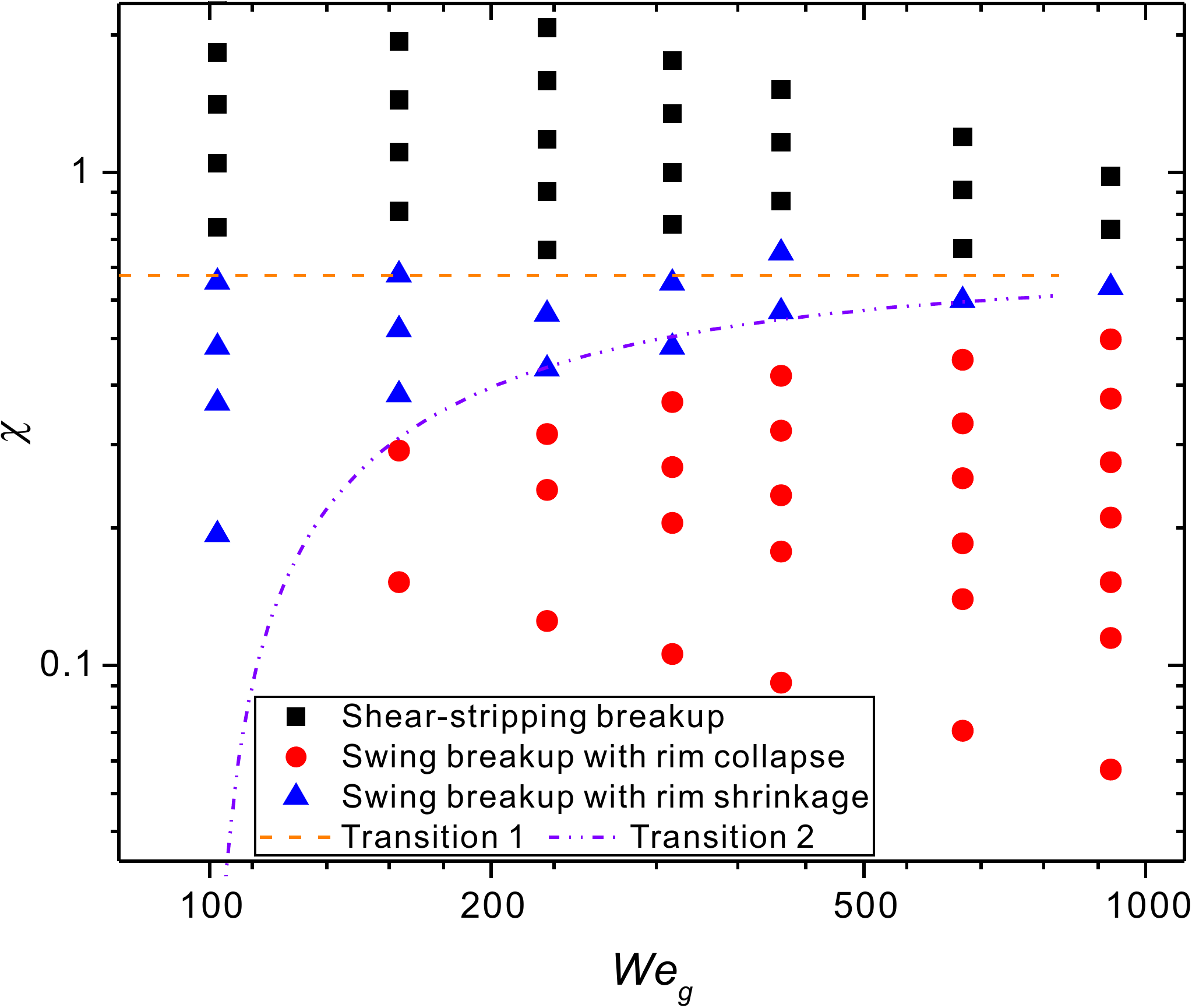}}
  \caption{Regime map of droplet breakup in airflow with a strong shear effect. Transition~1 is based on (\ref{eq:eq15}) and hence $\chi = 0.618$. Transition 2 is based on (\ref{eq:eq16}) and $ {{ \We}_{g}}-100= {{ \We}_{g}} ( {{\chi }^2}+\chi )$.}
\label{fig:fig10}
\end{figure}

To analyse further the shear effect of the airflow on droplet breakup, a regime map is produced as shown in figure \ref{fig:fig10}. According to the morphology of the droplet and the mechanism of breakup, the breakup of droplets can be categorised into three regimes, namely the swing breakup with front rim collapse (figures \ref{fig:fig04} and \ref{fig:fig18}), the swing breakup with front rim shrinkage (figure \ref{fig:fig17}), and the shear-stripping breakup (figure \ref{fig:fig19}). The shear-stripping breakup is the breakup regime of droplets in almost uniform airflow, and liquid sheets are stripped out continuously from the droplet body. In contrast, in a strong shear flow of air, the droplet body swings with a front rim, and the breakup can be divided into two regimes according to the collapse or the shrinkage of the front rim, as discussed in Section \ref{sec:sec32}.

The regime map is plotted in the parameter space of $\chi$ and $\We_g$. Here, $\chi$ reflects the relative effect of the droplet falling inertia, and together with $\We_g$, $\chi$ further affects the deformation of the droplet body during the interaction of the droplet with the shear layer, as discussed in Section \ref{sec:sec33}. It should be noted that the jet velocity and the shear strength of the jet shear layer are related to each other through the thickness of the shear layer (i.e., ${{u}_{g}}=\dot{\gamma }\delta $). Since the thickness of the shear layer $\delta$ almost does not vary with the jet velocity (as shown in figure \ref{fig:fig03}), the gas Weber number ($\We_g$) can represent not only the strength of aerodynamic force but also the strength of the shear effect. The change in the temporal evolution of the droplet with different parameters can be attributed to the variation in the relative strength at different stages. With the increase in $\chi$ under the same $\We_g$ (figures \ref{fig:fig04}, \ref{fig:fig17}, and \ref{fig:fig19}), the thickness of the early stretched sheet at the tail of the droplet is almost unchanged due to the same $\We_g$, but the droplet body deforms faster. Hence the swing breakup after the deformation of the droplet body consumes more liquid and its strength is greater, and the front rim contains less liquid. The breakup regime changes gradually from the swing breakup with rim shrinkage to the swing breakup with rim collapse, further to the shear--striping breakup in which the front rim disappears. With the increase in $\We_g$ under the same $\chi$ (figures \ref{fig:fig04} and \ref{fig:fig18}), the droplet body deforms faster, but the droplet breaks up faster at different stages. At large $\We_g$, small-scale breakup occurs continuously from the droplet tail, and the distinction between different stages becomes less clear.

To distinguish droplet breakup in shear flow from that in uniform flow, we compare the characteristic time for the droplet passing through the shear layer ($t_y={\left( {d_0}+\delta  \right)}/{{{u}_{d}}}$) with the time scale of droplet deformation (${{t}^{*}}={{{d}_{0}}}/({{u_g}\sqrt{{{{\rho }_g}}/{{{\rho }_d}}}})$), and find that ${t_y}/{t^{*}}\propto \chi $. When $\chi$ is small, the residence time scale (${{{t}_{y}}}/{{{t}^{*}}}$) of the droplet in the shear layer during the process of droplet deformation is small, and the droplet can pass quickly through the shear flow. In this condition, the droplet does not form a rim in the front of the droplet, but is stretched directly to form a liquid sheet around the edge of the windward surface, which is the same as the shear-stripping breakup in uniform airflow. In contrast, when $\chi$ is large, the droplet deforms mainly in the shear flow. The droplet will form a front rim, and the main body of the droplet will exhibit swing breakup under the strong shear effect. Therefore, the transition from the swing breakup to the shear-stripping breakup occurs when $\chi$ is larger than a threshold. The transition condition can be estimated from
\begin{equation}\label{eq:eq15}
  	{{ \We}_{f}}={{\We}_{g}},
\end{equation}
which corresponds to the transition condition of the droplet body deformation from being controlled by the shear layer uplift and the droplet falling inertia (i.e., the swing breakup) to being controlled by the aerodynamic force of a uniform airflow (i.e., the shear-stripping breakup). From (\ref{eq:eq6}) and (\ref{eq:eq15}), we can obtain the value of $\chi$ as $\chi = 0.618$. This transition condition is shown as Transition 1 in figure \ref{fig:fig10}, which is in good agreement with the experimental data.

For the droplet breakup under the strong shear effect, whether the front rim shrinks or collapses depends on the strength of the swing breakup, that is, the relative amount of liquid contained in the front rim and the swinging liquid film. The amount of liquid consumed in the swing breakup is less when the swing breakup develops quickly, and the rim is thicker when the droplet body deforms slowly. Here, we compare the relative speed of the swing breakup (i.e., $ {{ \We}_{g}}- {{ \We}_{g,c}}$, where $\We_{g,c}$ is the critical Weber number for sheet stretching and $\We_{g,c} = 100$ \citep{Guildenbecher2009SecondaryAtomization, Jackiw2021InternalFlow}) and the deformation of the droplet body (i.e., $\We_f$, which can be obtained from (\ref{eq:eq6})) after the tail sheet is stretched out, so we can obtain the transition condition of the front rim from collapse to shrinkage:
\begin{equation}\label{eq:eq16}
  	{{ \We}_{g}}- {{ \We}_{g,c}}={{\We}_{f}}.
\end{equation}
As shown by Transition 2 in figure \ref{fig:fig10}, the transition condition of the front rim from collapse to shrinkage based on (\ref{eq:eq16}) agrees well with the experimental data.

\subsection{Stretching of liquid film}\label{sec:sec35}
Different from the rapid breakup of the liquid film stretched from the droplet edge in a uniform airflow, the stretching of the liquid film in shear airflow is significant, and has an important influence on the uplift of the droplet front rim and the size of the fragments. We extract the boundary of the liquid film from the bottom-view images via image analysis, and obtain the temporal evolution of the liquid film area during the droplet deformation and breakup process.

\begin{figure}
  \centerline{\includegraphics[width=0.8\columnwidth]{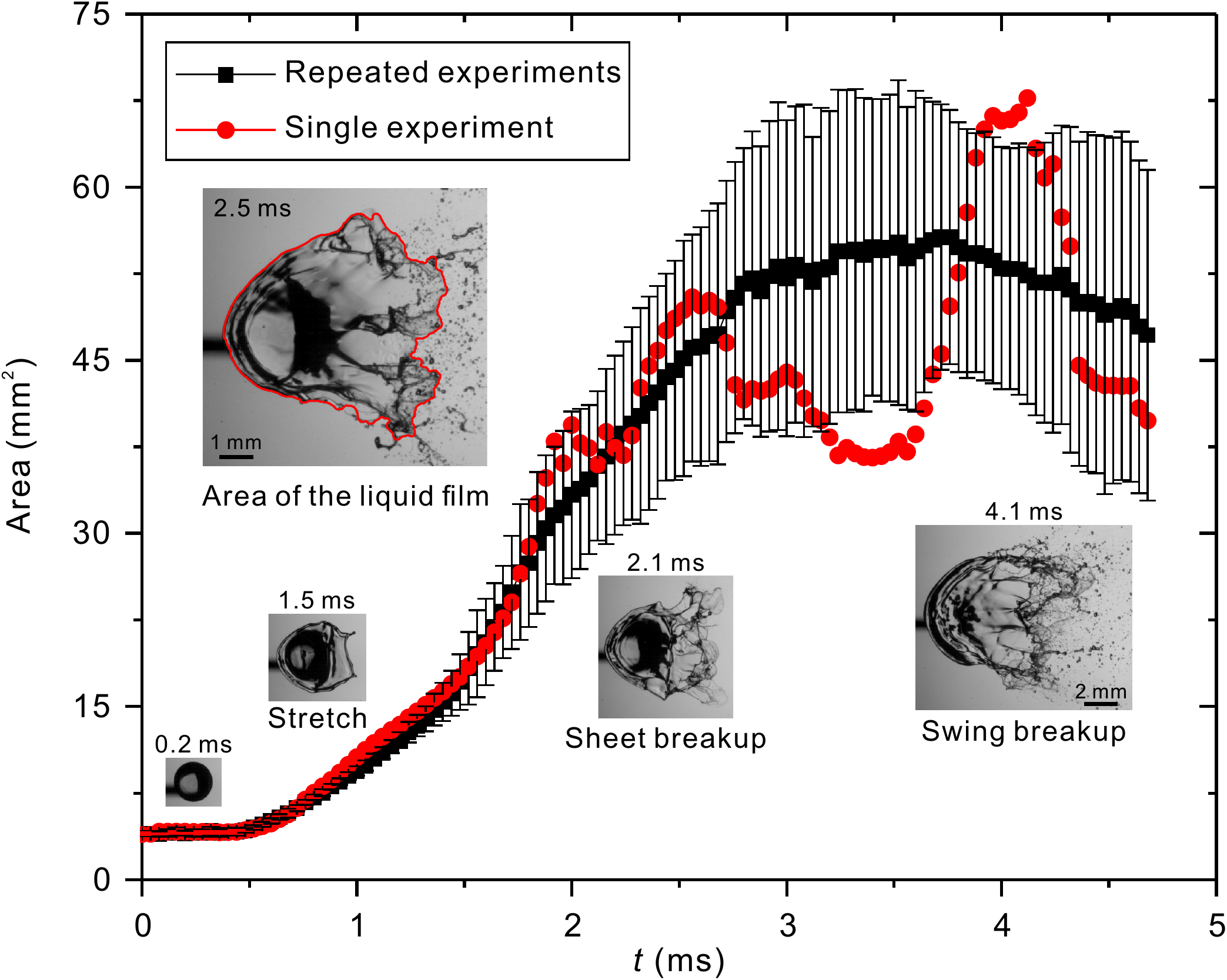}}
  \caption{Area evolution of the liquid film from the instant when the droplet edge just enters the shear flow to the end of the swing breakup stage. Here, $\We_g = 230$, $\We_d = 19.9$, $\chi = 0.295$, ethanol droplet with $d_0 = 2.4$ mm. The red curve shows the results of a single experiment, and the black curve shows the statistic results from repeated experiments. The error bars indicate the standard deviations of 20 repeated experiments.}
\label{fig:fig11}
\end{figure}

Figure \ref{fig:fig11} shows the evolution of the liquid film area from the instant when the droplet edge just enters the shear flow to the end of the swing breakup stage. From the development of the liquid film in a single experiment, it can be seen that the liquid film stretches out from the tail of the droplet at first, and its area increases gradually until it breaks up. Every sudden decrease in the area of the liquid film corresponds to the occurrence of breakup. At the instant of the film collapsing during the swing breakup stage, the area of the liquid film suddenly decreases from its maximum and does not increase any more. Since the development of instability on the liquid film is extremely quick and the liquid film after stretching is affected easily by turbulence, there is a strong uncertainty about the moment of the liquid film breakup. To compare the results under different parameters, we repeat a large number of experiments to get a statistical result as shown in figure \ref{fig:fig11}.

\begin{figure}
  \centerline{\includegraphics[width=1\columnwidth]{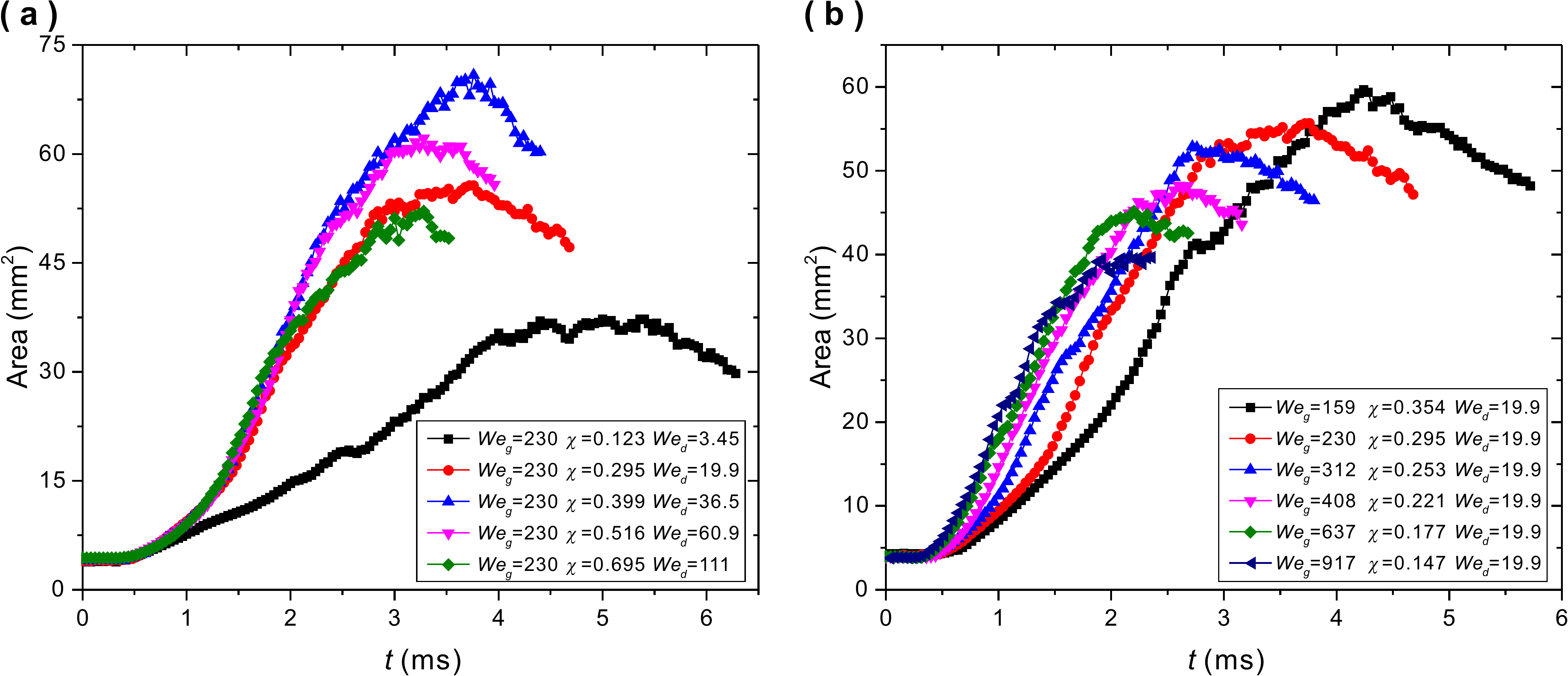}}
  \caption{Temporal evolution of the area of the liquid film under the influence of (a) $\chi$ and (b) $\We_g$. Each curve shows the average of 20 repeated experiments.}
\label{fig:fig12}
\end{figure}

The temporal evolution of the liquid film area at different $\chi$ is shown in figure \ref{fig:fig12}a. In the case of $\We_g = 230$, as $\chi$ increases from 0.123 to 0.695 (corresponding to $\We_d$ increasing from 3.45 to 111), the stretch speed of the early liquid film tends to be the same, the average time to complete swing breakup becomes shorter, and the maximum liquid film area first increases and then decreases. When $\chi$ is very small ($\chi = 0.123$), the early liquid sheet at the tail of the droplet deforms slowly and contains much liquid in the low-speed region of the airflow, which leads to a slow development speed and a long development time of the film area at the early stage. After the sheet breakup stage, the liquid left for the swing breakup is little, and the swing breakup is weak, which results in a small maximum area of the liquid film. As $\chi$ increases from 0.123 to 0.399, the downward inertia of the droplet increases, and more liquid quickly invades the shear flow. The early sheet is stretched quickly in the high-speed region of the airflow, and the stretch speed tends to be the same because $\We_g$ is constant. Combined with the uplift effect of the shear layer, the increase of $\chi$ also leads to a faster deformation of the droplet main body, an increased windward area of the droplet, more liquid contained in the liquid tongue before the swing breakup, and finally more intensive swing breakup. Therefore, the maximum area of the liquid film increases as $\chi$ increases. If $\chi$ increases further ($\chi > 0.399$), then the stretch speed of the early sheet remains unchanged, but the maximum area of the liquid film decreases with increasing $\chi$. This is because the further increase in $\chi$ results in too little liquid in the front rim of the droplet. Since the liquid in the front rim is insufficient, when the swing breakup occurs, the front rim cannot maintain its original position but moves downstream, which weakens the stretching of the liquid film. In addition, as $\chi$ increases from 0.123 to 0.695, the average time to the end of swing breakup becomes shorter because the droplet enters the shear flow faster to deform.

Further, combined with the regime map of figure \ref{fig:fig10}, the area evolution of the liquid film with different $\chi$ (i.e., varied $\We_d$ with the fixed $\We_g$) in figure \ref{fig:fig12}a can be linked with the breakup regimes. As $\chi$ increases from 0.123 to 0.399, the maximum area of the liquid film increases. This is because the swing breakup strength at the tail of the droplet body increases while the front rim remains relatively stable, which strengthens the stretching of the liquid film. This means that the swing breakup at the tail cannot make the front rim move, and after the swing breakup, the front rim enters the high-speed region of the airflow and then collapses, corresponding to the regime of the swing breakup with rim collapse. As $\chi$ increases from 0.399 to 0.695, the swing breakup strength at the tail of the droplet body is further enhanced, but the front rim is affected by the swing breakup and cannot maintain its position, which weakens the stretching of the liquid film and finally causes a reduction in the maximum area of the liquid film. In this condition, the rim will be taken away from the main stream by the swing breakup and then shrinkage, corresponding to the regime of the swing breakup with rim shrink. Finally, when $\chi = 0.695$, the front rim disappears, and the droplet body is pierced after the early stripping along the droplet periphery, which causes the liquid film to disappear quickly after the early sheet stretching, corresponding to the regime of shear-stripping breakup in the uniform airflow. These results reflect the connection between the swing breakup strength and the movement of the front rim, which determines the regime of breakup.

The temporal evolution of the area of the liquid film at different $\We_g$ is shown in figure \ref{fig:fig12}b. In the case of $\We_d = 19.9$, as $\We_g$ increases from 159 to 917 and $\chi$ decreases from 0.354 to 0.147, the stretching speed of the early liquid film becomes faster, the average time to complete the swing backup becomes shorter, and the maximum liquid film area becomes smaller. This is because an increase in $\We_g$ means a stronger shear effect, and the droplet will stretch and break up faster. And due to the rapid and continuous occurrence of sheet breakup caused by the stronger shear effect, the swing breakup becomes weaker, and the maximum area of the liquid film becomes smaller.

\subsection{Fragment size distribution}\label{sec:sec36}
The fragment size after droplet breakup is the main concern in many practical applications. Under our experimental conditions with high $\We_g$ accompanied by strong shear, the fragment size is analysed from the high-speed images. Due to the long duration of the whole breakup process, the fragments produced in different stages after the completion of the breakup are already dispersed to a large region. In the traditional method of extracting the size of fragments from a single image, it will require a large shooting area to cover all the fragments, which will mean that most small fragments cannot be counted under the limitation of spatial resolution. To circumvent the tradeoff between the shooting area and the spatial resolution, we develop a customised Matlab program based on fragment tracking on image sequences. By tracking each fragment, we can obtain the size of all fragments after the completion of the breakup using only a small shooting area. By tuning the photography setting and image-processing procedure, we can obtain all the fragments size larger than 100 $\mu$m. A detailed description of the procedure for measuring the fragment size is given in the supplementary material.

Figure \ref{fig:fig13} shows the probability density of fragment size weighted by volume, which represents the ratio of the total volume of fragments with a certain diameter to the fragments' total volume.
Under the same $\We_g$ ($\We_g = 230$), the different $\chi$ in figure \ref{fig:fig13} correspond to the three breakup regimes, namely the swing breakup with rim collapse ($\chi = 0.123$), the swing breakup with rim shrinkage ($\chi = 0.399$), and the shear-stripping breakup ($\chi = 1.535$), respectively. Due to the differences in the breakup mechanism and the breakup morphology, the fragment size distribution is also different.
For the shear-stripping breakup ($\chi = 1.535$), the distribution appears unimodal with a peak at $d_p \sim 0.17$ mm, which is similar to the result of Guildenbecher \emph{et al.} \citeyearpar{Guildenbecher2017FragmentSize} in uniform airflow. For the swing breakup with rim collapse ($\chi = 0.123$), the size distribution also appears unimodal.
But compared with the shear-stripping breakup, the swing breakup with rim collapse generally occurs in a region where the airflow velocity is lower. In addition, the droplet body is almost parallel to the airflow when the front rim collapses. Hence the windward area is small, and less energy is transferred from the airflow to the droplet during the breakup process. The lower local air velocity and less energy transfer leads to larger fragments during the swing breakup with rim collapse than the shear-stripping breakup. For swing breakup with rim shrinkage ($\chi = 0.399$), the front rim is driven out of the shear layer by the strong swing effect and then shrinks, producing larger fragments. We calculate the characteristic diameters of the fragments after droplet breakup in table \ref{tab:tab2}. At $\We_g = 230$ and as $\chi$ increases from 0.123 to 1.535, the size distribution first widens and then narrows. Meanwhile, the characteristic diameters, including the Sauter mean diameter (SMD), the mass median diameter (MMD) and the peak diameter ($d_p$), increase first and then decrease.

\begin{figure}
  \centerline{\includegraphics[width=1\columnwidth]{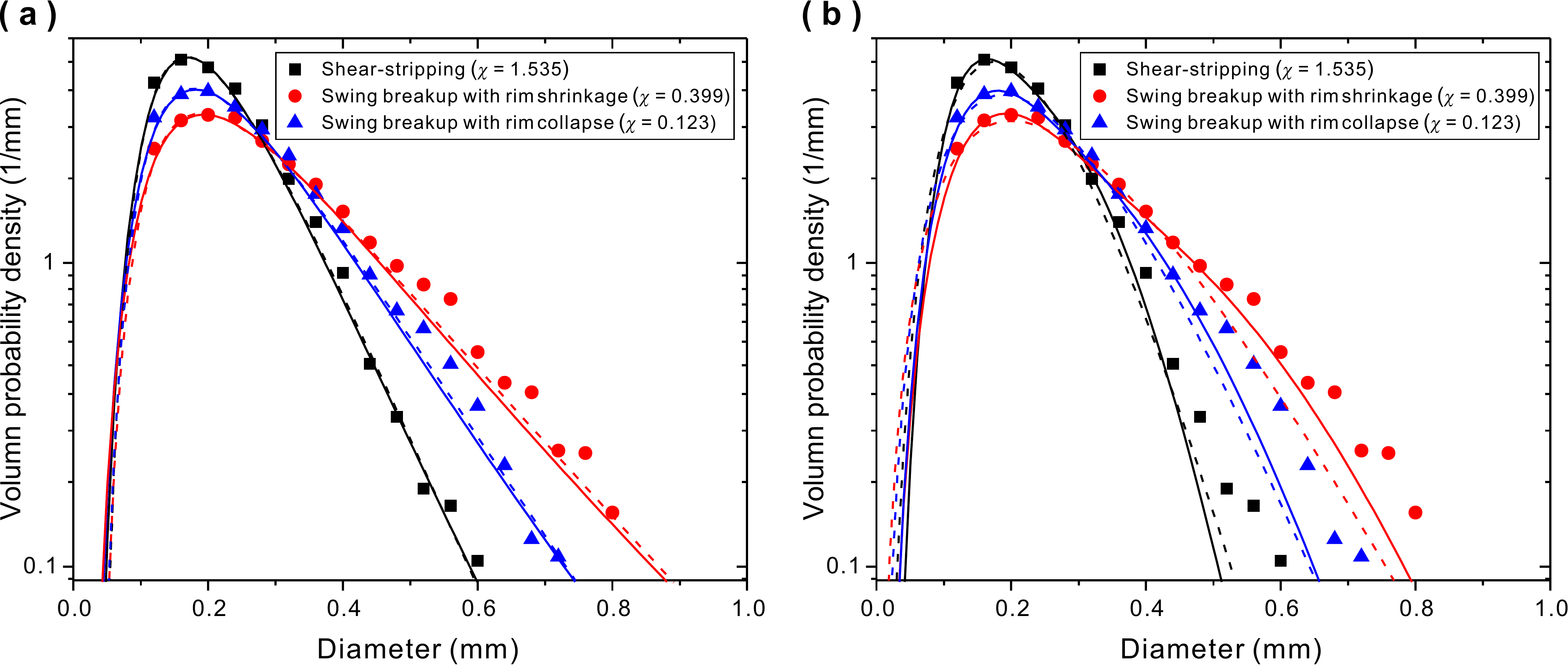}}
  \caption{Volume probability density distribution of the fragments after completion of the droplet breakup. (a) The solid lines are the best-fit log-normal distributions, and the dashed lines are the best-fit infinitely divisible distributions. (b) The solid lines are the best-fit compound gamma distributions, and the dashed lines are the best-fit gamma distributions. Each scatterplot shows the statistical results of about 30 repeated experiments under the same conditions. Here, $\We_g = 230$.}
\label{fig:fig13}
\end{figure}

\begin{table}
\begin{center}
\def~{\hphantom{0}}
\begin{tabular}{p{2.5cm}<{\centering}p{2.5cm}<{\centering}p{2.5cm}<{\centering}p{2.5cm}<{\centering}}
  $\chi$		& SMD (mm)		& MMD (mm)		& $d_p$ (mm) \\
  0.123		& 0.228		& 0.250		& 0.19  \\
  0.399		& 0.256		& 0.287		& 0.20 \\
  1.535		& 0.182		& 0.219		& 0.17\\
\end{tabular}
\caption{Characteristic diameters of the fragments after droplet breakup.}
\label{tab:tab2}
\end{center}
\end{table}

In the spray community, there are many functions to describe the fragment size distribution. Among them, the log-normal distribution is used widely, which is derived from considering the overall breakup process as sequential cascades where liquid volumes are aggregated or split at random \citep{Gorokhovski2003Distribution, HadjAchour2021Distribution, Kolmogorov1941Lognormal}. The log-normal distribution is
\begin{equation}\label{eq:eq17}
  {{f}_{\log }}\left( d \right)=\frac{1}{\sqrt{2\pi }d{{\sigma }_{\log }}}\exp \left[ -\frac{1}{2\sigma _{\log }^{2}}{{\left( \ln d-\ln {{d}_{m}} \right)}^{2}} \right],
\end{equation}
where $d_m$ is the geometric mass median diameter, and $\sigma_{\log}$ is a scaling parameter.

Based on sequential cascades of breakups, there are many variants of the log-normal distribution \citep{Dommermuth1997Distribution, Rimbert2004Logstable, Gorokhovski2012Distribution}. More recently, the infinitely divisible distribution proposed by Novikov and Dommermuth \citeyearpar{Dommermuth1997Distribution} was used by Vallon \emph{et al.}\ \citeyearpar{Romain2021Distribution}, which assumed that the liquid underwent a sequential cascading breakup in turbulent sprays similar to the intermittent turbulent dissipation. The infinitely divisible distribution is
\begin{equation}\label{eq:eq18}
  {{f}_{\inf }}\left( d \right)=\frac{d_{\inf }^{3/2}}{\sqrt{2\pi }{{\sigma }_{\inf }}{{d}^{3/2}}}\exp \left[ -\frac{{{d}_{\inf }}}{2\sigma _{\inf }^{2}}{{\left( {{d}_{\inf }}{{d}^{-1/2}}-{{d}^{1/2}} \right)}^{2}} \right],\quad d\ge 0,
\end{equation}
where $d_{\inf}$ and $\sigma_{\inf}$ are the average and the standard deviation, respectively.

The fitting results of (\ref{eq:eq17}) and (\ref{eq:eq18}) for different breakup modes are shown in figure \ref{fig:fig13}a. The log-normal distribution and the infinitely divisible distribution are very close, and both of them can describe the fragment size distribution reasonably well, especially for the shear-stripping breakup. Due to the high $\Re$ of airflow and the high $\We_g$, the strong turbulence in the shear layer of airflow masks the effect of different breakup mechanisms on the final fragment size distribution to some extent. Therefore, the size distribution is similar to the result of a random cascade breakup, and the different breakup regimes affect only the mean and variance of the distribution. But for the droplet breakup in the shear layer, there are some deviations at diameter $\sim$0.6 mm, especially for the swing breakup with rim shrinkage.

In addition, the gamma distribution is known to be a good approximation of the log-normal distribution \citep{HadjAchour2021Distribution} and was used recently by Vallon \emph{et al.}\ \citeyearpar{Romain2021Distribution} and Huck \emph{et al.}\ \citeyearpar{Aliseda2022CoaxialJet}. Different from the log-normal distribution and infinitely divisible distribution, the gamma distribution is based on the assumption that the fragments come from the rupture of objects in the form of threads or ligaments, i.e., the elementary distribution of the fragment size originates from the ligament breakup \citep{Villermaux2020FragmentationCohesion, Villermaux2011Distribution, Villermaux2004Gamma}. The gamma distribution is
\begin{equation}\label{eq:eq19}
  	g\left( d \right)=\frac{{{d}^{\alpha -1}}{{e}^{-\frac{d}{\beta }}}}{{{\beta }^{\alpha }}\Gamma \left( \alpha  \right)},
\end{equation}
where $\alpha$ is the shape parameter and refers to the roughness of the ligament, and $\beta$ is the scale parameter. The fitting results of (\ref{eq:eq19}) for different breakup modes are shown in figure \ref{fig:fig13}b. The gamma distribution displays an exponential fall-off at the tail and a relatively narrow distribution. For the droplet breakup in the shear layer of airflow, the gamma fitting underestimates the experimental data at the large fragment sizes.

Since the fragments are formed mainly in two different stages, and the single gamma distribution displays an apparent discrepancy at the large fragment sizes, we further use the compound gamma distribution proposed by Villermaux \emph{et al.}\ \citeyearpar{Villermaux2020FragmentationCohesion, Villermaux2011Distribution} to describe the fragment size distribution. The compound gamma distribution is based on the overall fragments originating from two distinct classes of ligaments. The linear superposition of the two elementary gamma distributions leads to a compound distribution
\begin{equation}\label{eq:eq20}
  	{{g}_{c}}\left( d \right)={{A}_{1}}{{g}_{1}}\left( d \right)+{{A}_{2}}{{g}_{2}}\left( d \right),
\end{equation}
where $g_1(d)$ and $g_2(d)$ are the elementary gamma distributions with different ligament roughnesses, and $A_1$ and $A_2$ are the weights of $g_1(d)$ and $g_2(d)$, with $A_1+A_2=1$. The fitting results of (\ref{eq:eq20}) for different breakup modes are shown in figure \ref{fig:fig13}b. The compound gamma distribution shows clearly a better description of the experimental data only for the swing breakup with rim shrinkage. This is due to the apparent difference in the sizes of fragments produced by the swing breakup of the droplet body and the shrinkage of the front rim. But for the swing breakup with rim collapse, because the sizes of fragments produced in different stages are close, the elementary gamma distributions are difficult to distinguish from each other. Hence the parameters in the compound gamma distribution are not stable, and the compound distribution has no obvious superiority in performance.

\begin{figure}
  \centerline{\includegraphics[width=0.6\columnwidth]{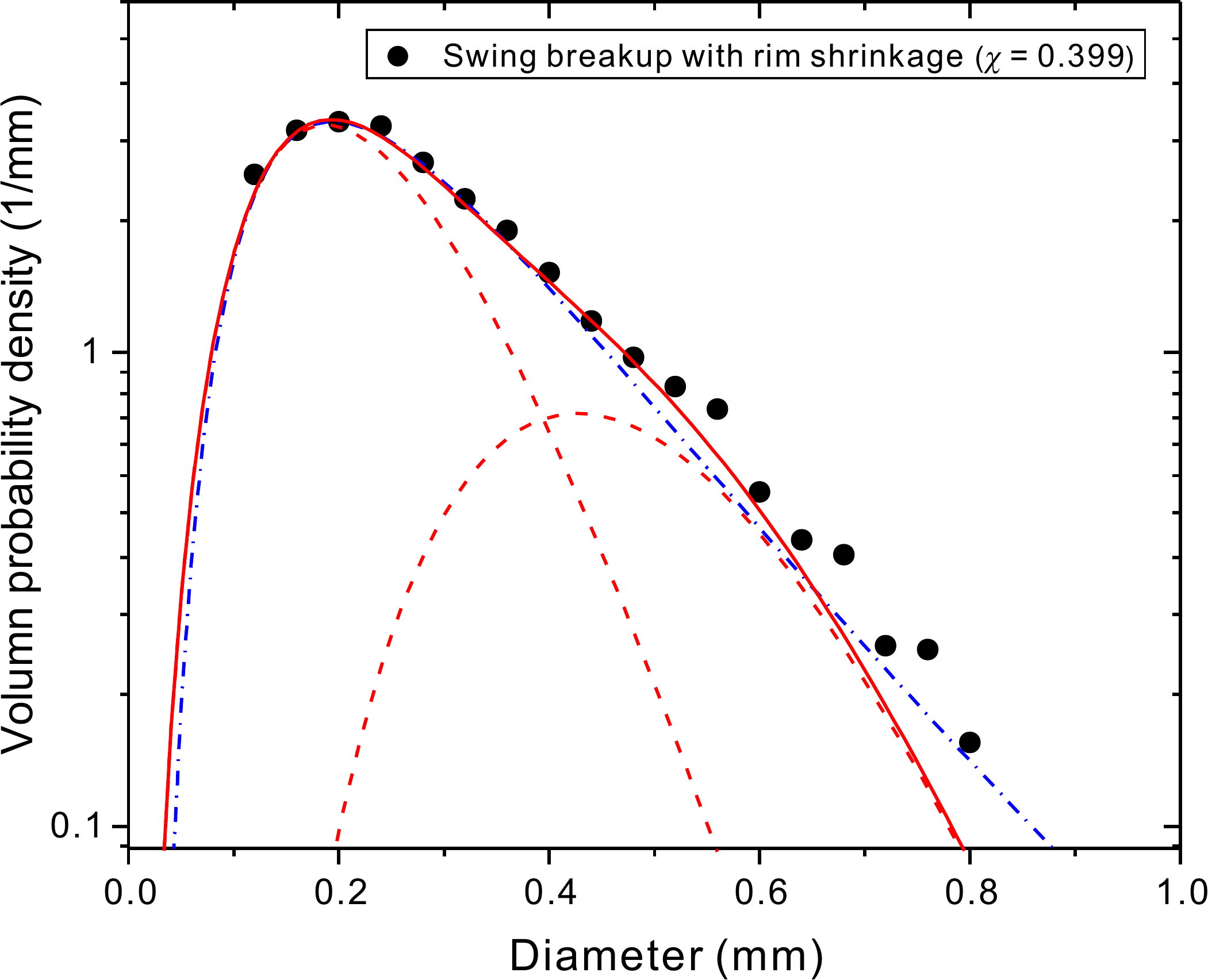}}
  \caption{Volume probability density distribution of the fragments after completion of the droplet breakup. The blue dashed-dotted line is the best-fit log-normal distribution, the red solid line is the best-fit compound gamma distribution, and the red dashed lines are the sub-gamma distributions. Each scatterplot shows the statistical results of about 30 repeated experiments under the same conditions. Here, $\We_g = 230$.}
\label{fig:fig14}
\end{figure}

A comparison between the log-normal distribution and the compound gamma distribution for the swing breakup with rim shrinkage is shown in figure \ref{fig:fig14}. The log-normal distribution seems to be slightly better across the entire size range, but the compound gamma distribution is superior at the intermediate size and reflects the effects of the two different fragmentation mechanisms on the size distribution.

Finally, table \ref{tab:tab3} presents the mean and variance obtained by fitting different distributions, and compares them with the experimental results; and the calculation of the mean and variance is given in Appendix \ref{sec:AppC}. The log-normal distribution and infinitely divisible distribution can fit the mean of the experimental data better, while the gamma and compound gamma distributions can fit the variance better. This is consistent with the view of Villermaux \emph{et al.}\ \citeyearpar{Villermaux2007Fragmentation, Villermaux2020FragmentationCohesion} that the gamma distribution can model the experimental data well with prior knowledge of the mean size.

\begin{table}
\begin{center}
\def~{\hphantom{0}}
\begin{tabular}{p{0.8cm}<{\centering}p{1cm}<{\centering}p{1cm}<{\centering}p{1cm}<{\centering}p{1cm}<{\centering}p{1cm}<{\centering}p{1cm}<{\centering}p{1cm}<{\centering}p{1cm}<{\centering}p{1cm}<{\centering}p{1cm}<{\centering}}
\multirow{2}{*}{$\chi$} & \multicolumn{2}{c}{Experimental} & \multicolumn{2}{c}{Log-normal} & \multicolumn{2}{c}{Infinite-divisible} & \multicolumn{2}{c}{Gamma} & \multicolumn{2}{c}{Compound Gamma} \\
                        & $\EX_{ex}$           & $\DX_{ex}$            & $\EX_{\log}$         & $\DX_{\log}$          & $\EX_{\inf}$             & $\DX_{\inf}$              & $\EX_{g}$         & $\DX_{g}$         & $\EX_{c}$             & $\DX_c$              \\
0.123                   & 0.275          & 0.0167          & 0.271         & 0.0224         & 0.272             & 0.0215             & 0.250       & 0.0157      & 0.259           & 0.0159           \\
0.399                   & 0.314          & 0.0247          & 0.315         & 0.0379         & 0.318             & 0.0362             & 0.286       & 0.0245      & 0.299           & 0.0256           \\
1.535                   & 0.234          & 0.0094          & 0.229         & 0.0116         & 0.230             & 0.0112             & 0.216       & 0.0090      & 0.223           & 0.0086
\end{tabular}
\caption{Mean and variance of experimental results and the fitting distributions}
\label{tab:tab3}
\end{center}
\end{table}

\subsection{Effect of the Ohnesorge number}\label{sec:sec37}
\begin{figure}
  \centerline{\includegraphics[width=0.7\columnwidth]{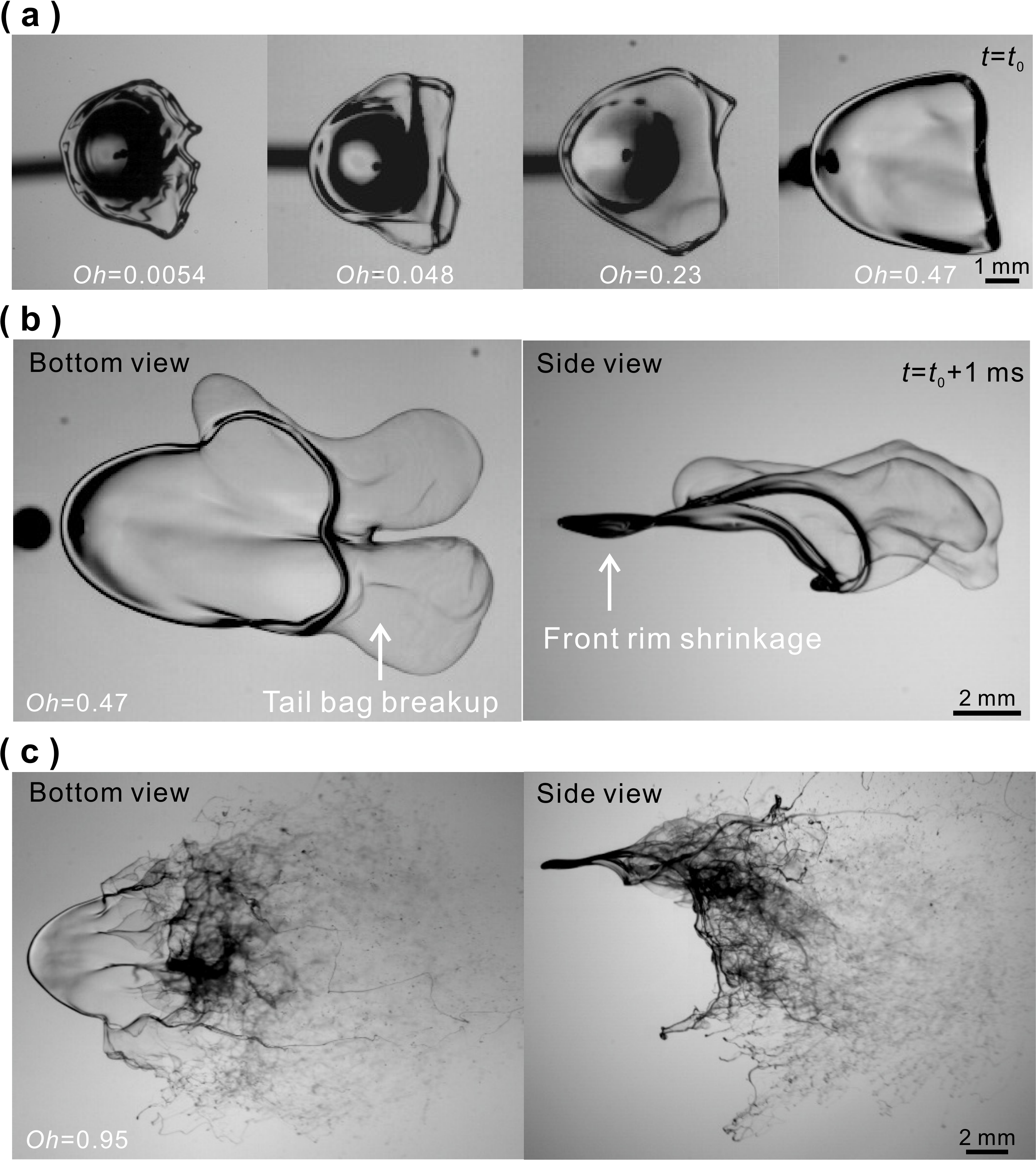}}
  \caption{(a) Bottom-view images of the tail sheet before its first breakup at different $\Oh$. The values of $\Oh$ are, respectively, 0.0054, 0.048, 0.23, 0.47, with corresponding $\We_g$ = 230, 239, 230, 219, $\chi$ = 0.294, 0.347, 0.353, 0.354, and $d_0$ = 2.4, 2.3, 2.3, 2.2 mm. (b)~Synchronised images of the bottom and side views after tail bag development at $\Oh$ = 0.47, $\We_g$ = 219, $\chi$ = 0.354, silicone oil with $d_0$ = 2.2 mm (movie 5 in the supplementary material). (c) Synchronised images of the bottom and side views for the swing breakup of the droplet body at $\Oh$ = 0.95, $\We_g = 873$, $\chi = 0.177$ (movie 6 in the supplementary material).}
\label{fig:fig15}
\end{figure}

The Ohnesorge number $\Oh$ indicates the ratio of droplet viscous forces to surface tension forces. In the process of droplet breakup, the droplet viscosity induces energy dissipation and hinders droplet deformation. Hence $\Oh$ is an important control parameter for the droplet breakup. Instead of using ethanol in Sections \ref{sec:sec32}--\ref{sec:sec36}, we use silicone oil of different viscosities to study the effect of $\Oh$ on droplet breakup in strong shear airflow. Figure \ref{fig:fig15}a shows images of the tail sheet before its first breakup at different $\Oh$ with almost the same $\chi$ and $\We_g$. (the slight variations in $\chi$ and $\We_g$ are due to the variations in the fluid density and the surface tension as the viscosity changes, shown in table \ref{tab:tab1}.) With increasing $\Oh$, the stretched sheet at the tail is more stable, and the stretched area increases slightly, similar to the decrease in $\We_g$ as discussed in Section \ref{sec:sec32} and figure \ref{fig:fig05}. When $\Oh$ increases to 0.47, the stretched sheet transforms to a tail bag structure due to the stabilisation effect of the high viscosity, as shown in figure \ref{fig:fig15}b. This process is similar to the transition from shear-stripping breakup to multimode breakup in uniform airflow when increasing $\Oh$ \citep{Guildenbecher2009SecondaryAtomization, Radhakrishna2021HighOh}. In this condition, because the rapid development of the tail bag consumes much liquid, the swing breakup of the droplet body does not occur, and the front rim is lifted and then shrinks. Although the increase in $\Oh$ inhibits the droplet deformation and even alters the breakup mode, the swing breakup of the droplet body still occurs if $\We_g$ is large enough. For example, even when $\Oh$ reaches 0.96 and $\We_g$ increases to 853, the droplet body swings and breaks up, as shown in figure \ref{fig:fig15}c.

For viscous droplets, the swing breakup also originates from the flattening deformation of the droplet body, but the flattened thickness of the droplet body and the development of the RT instability are both affected by the droplet viscosity. For the cases presented in this section, $100< {{ \We}_{f}}<490$, $ {{ \Re}_{f}}<100$ and $P>4.9$, which indicates the viscous case following the droplet collision analogue \citep{Roisman2009DeformationViscous}. In this condition, the flattened thickness of the droplet body is determined mainly by the flattening Reynolds number ($\Re_f$):
\begin{equation}\label{eq:eq21}
  \frac{{{h}_{v}}}{{{d}_{0}}}\sim \Re_{f}^{-2/5},
\end{equation}
where the subscript $v$ in $h_v$ indicates the viscous fluid, and $\Re_f$ indicates the ratio of the flattening force to the viscous force of the droplet, $ {{ \Re}_{f}}={{\rho }_{d}}{{d}_{0}}{{u}_{f}}/{{\mu }_{d}}$; $u_f$ can be obtained from (\ref{eq:eq7}). Hence $\Re_f$ can be transformed further into
\begin{equation}\label{eq:eq22}
  {{ \Re}_{f}}=\We_{g}^{1/2}\left({\chi ^2+\chi }\right)^{1/2}/{\Oh}.
\end{equation}
By substituting (\ref{eq:eq22}) into (\ref{eq:eq21}), we can predict the dimensionless thickness of the droplet body ($H_v$) as
\begin{equation}\label{eq:eq23}
  {{H}_{v}}=\frac{{{h}_{v}}}{{{d}_{0}}}={{C}_{2}}\We_{g}^{-1/5}{{\Oh}^{2/5}}{{\left( {{\chi }^{2}}+\chi  \right)}^{-1/5}}.
\end{equation}

Considering the RT instability of viscous fluids, when $\rho_l \gg \rho_g$, $\mu_l \gg \mu_g$, the dispersion equation of the instability wave \citep{Funada2002RTinstability, Joseph1999DropBreakup} is
\begin{equation}\label{eq:eq24}
  {{n}_{\RT,v}}=-\frac{{{k}^{2}}{{\mu }_{d}}}{{{\rho }_{d}}}\pm \sqrt{\frac{{{k}^{4}}{{\mu }_{d}}^{2}}{{{\rho }_{d}}^{2}}+ak-\frac{{{k}^{3}}\sigma }{{{\rho }_{d}}}}.
\end{equation}
By assuming that the viscous and surface tension effects are additive to the leading order, Aliseda \emph{et al.}\ \citeyearpar{Aliseda2008RTinstability} obtained the form of the most-amplified wavelength as
\begin{equation}\label{eq:eq25}
  {{\lambda }_{\RT,v}}=2\pi \left[ {{\left( \frac{3\sigma }{{{\rho }_{d}}a} \right)}^{1/2}}+{{\left( \frac{\mu _{d}^{2}}{\rho _{d}^{2}a} \right)}^{1/3}} \right],
\end{equation}
where $a$ is obtained by (\ref{eq:eq11}), but replacing the thickness of the droplet body $h$ by $h_v$ in (\ref{eq:eq23}). Therefore, we can predict the most-amplified wavelength of the RT instability of viscous fluids as
\begin{equation}\label{eq:eq26}
  \frac{{{\lambda }_{\RT,v}}}{{{d}_{0}}}=2\pi \left( \sqrt{3}\We_{g}^{-1/2}H_{v}^{1/2}+ {{ \Oh}^{2/3}}\We_{g}^{-1/3}H_{v}^{1/3} \right),
\end{equation}
where $H_v$ is given in (\ref{eq:eq23}).

\begin{figure}
  \centerline{\includegraphics[width=0.6\columnwidth]{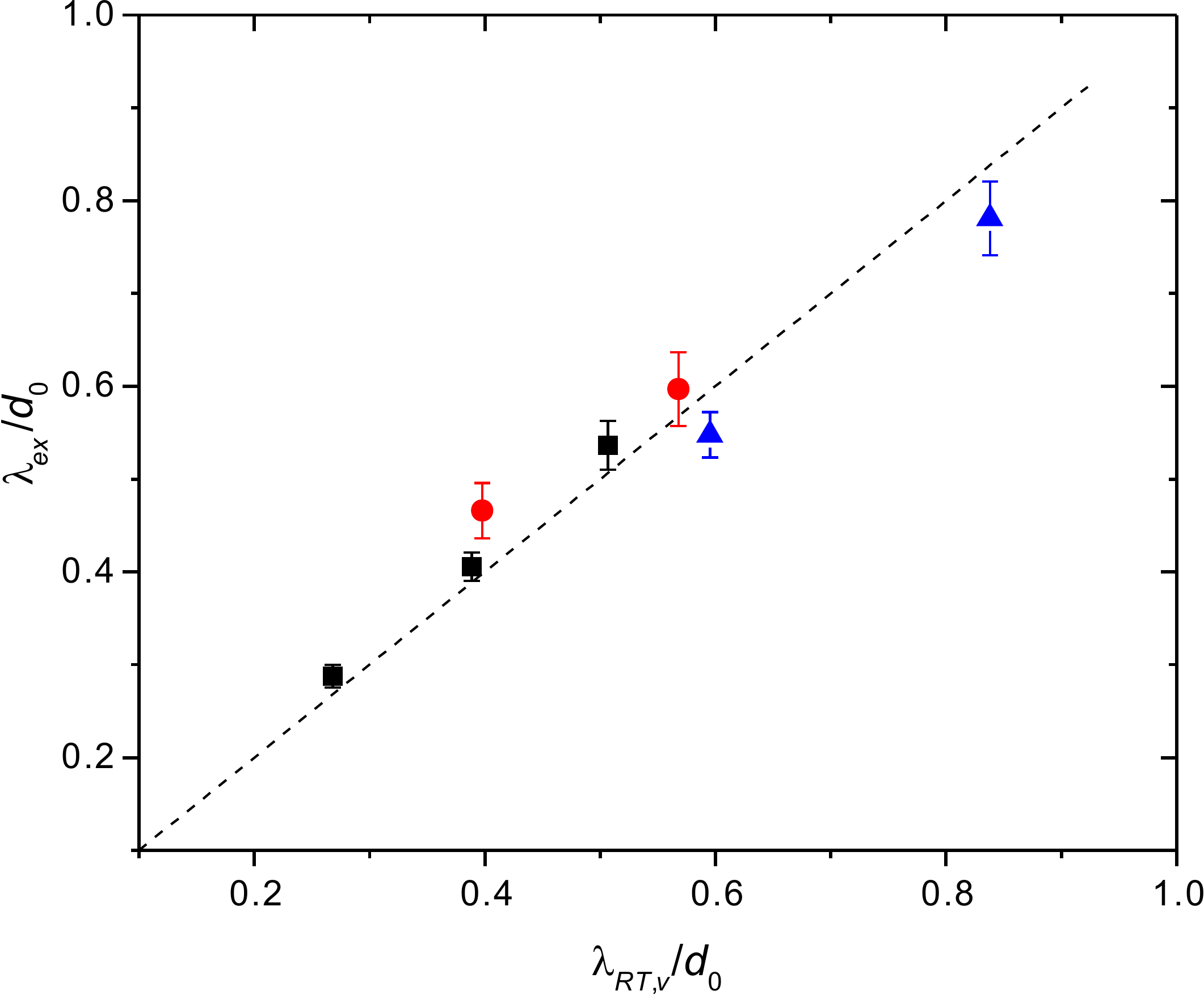}}
  \caption{Comparison between theoretical and experimental instability wavelengths, ${{\lambda }_{\RT,v}}$ and ${{\lambda }_{ex}}$, for viscous droplets. The experimental instability wavelength is obtained from experimental images, and the theoretical instability wavelength is obtained from (\ref{eq:eq26}) with fitted constant $C_2 = 0.75$. The dashed line indicates ${{\lambda }_{\RT,v}}={{\lambda }_{ex}}$. As the wavelength increases, the three black squares correspond to $\We_g =$ 921, 410, 230, respectively, $\chi =$ 0.163, 0.244, 0.326, respectively, and $\Oh =$ 0.23, silicone oil with $d_0 = 2.3$ mm. The two red circles correspond to $\We_g =$ 877, 390, respectively, $\chi =$ 0.163, 0.245, respectively, and $\Oh =$ 0.47, silicone oil with $d_0 =$ 2.2 mm. The two blue triangles correspond to $\We_g =$ 873, 388, respectively, $\chi = 0.164$, 0.245, respectively, and $\Oh = 0.95$, silicone oil with $d_0 = 2.2$ mm.}
\label{fig:fig16}
\end{figure}

The experimental wavelengths of the RT instability wave are also obtained from the high-speed images according to (\ref{eq:eq14}). A comparison between the theoretical and experimental wavelengths of the RT instability wave is made in figure \ref{fig:fig16}, which shows that the experimental wavelengths can be well predicted by (\ref{eq:eq26}) for high-viscosity fluids with fitting parameter $C_2 = 0.75$. The above analysis confirms that the droplet viscosity can affect the development of the RT instability in the swing breakup.

\section{Conclusions}\label{sec:sec4}
In this study, we investigate experimentally the strong shear effect on droplet breakup in airflows, in which the shear effect is generated by the shear layer of an air jet. Under a strong shear effect, the droplet breakup process is divided into three stages: the early sheet breakup stage that is featured by the sheet stretching at the droplet tail; the swing breakup stage that is featured by the intensive swing of the droplet body with asymmetric deformation; and the final rim breakup stage that is featured by the shrinkage or collapse of the rim at the droplet front. The thickness of the droplet body is predicted by analogising the flattening deformation of the droplet body in the shear flow to inertia-dominated droplet collisions and referring to the wing lift theory. Then the predicted thickness is used to calculate the most-amplified wavelength of the RT instability of the droplet body. The comparison of theoretical and experimental wavelengths reveals that the swing breakup is governed by the transverse RT instability based on the droplet deformation.

Different regimes are divided in the regime map of droplet breakup, and scaling analysis is performed for the transition boundaries between different regimes. The transition between the swing breakup regime in shear flow and the shear-stripping breakup regime in uniform flow is $ {{ \We}_{f}}= {{ \We}_{g}}$, and the transition between collapse and shrinkage of the front rim in the swing breakup regimes is $ {{ \We}_{g}}- {{ \We}_{g,c}}= {{ \We}_{f}}$. The quantitative analysis of the area of the stretched liquid film indicates that the stretching effect of shear airflow on the droplet is determined by the relative strength of different stages, and the large area of the stretched liquid film often corresponds to a strong swing breakup. The size distribution of the fragments after the droplet breakup is compared, and it is found that the front rim generated by the uplift effect of the shear airflow causes the fragment size distribution to be more dispersed. Different size distribution functions are compared. The log-normal distribution and the infinitely divisible distribution can describe the fragment size distribution across the entire size range. The compound gamma distribution is superior at the intermediate size and can reflect the effects of the two different fragmentation mechanisms on the size distribution. Finally, the viscosity of the droplet can inhibit the droplet breakup in a strong shear airflow, and droplets with a higher Ohnesorge number require a larger $\We_g$ to achieve similar breakup morphology. The theoretical analysis reveals that the deformation of the droplet body and the development of the RT instability are both affected by the droplet viscosity.

\newcounter{savesecnumdepth}
\providecommand{\backsectionname}{Acknowledgements}
\newcommand{\backsection}[2][\backsectionname]{\begingroup\par%
  \small%
  \setcounter{savesecnumdepth}{\value{secnumdepth}}%
  \setcounter{secnumdepth}{0}%
\vskip6pt
\noindent \textbf{#1.} #2\par%
  \setcounter{secnumdepth}{\value{savesecnumdepth}}%
  \endgroup}

\backsection[Supplementary data]{\label{SupMat}Supplementary material and movies are available online.}
\backsection[Acknowledgements]{This work was supported by the National Natural Science Foundation of China (Grant Nos. 51676137, 52176083, and 51921004). The authors are grateful to the anonymous reviewers for their valuable comments and suggestions.}
\backsection[Declaration of interests]{The authors report no conflict of interest.}

\appendix
\section{Image sequences in different conditions}\label{sec:AppA}
\begin{figure}
  \centerline{\includegraphics[width=1\columnwidth]{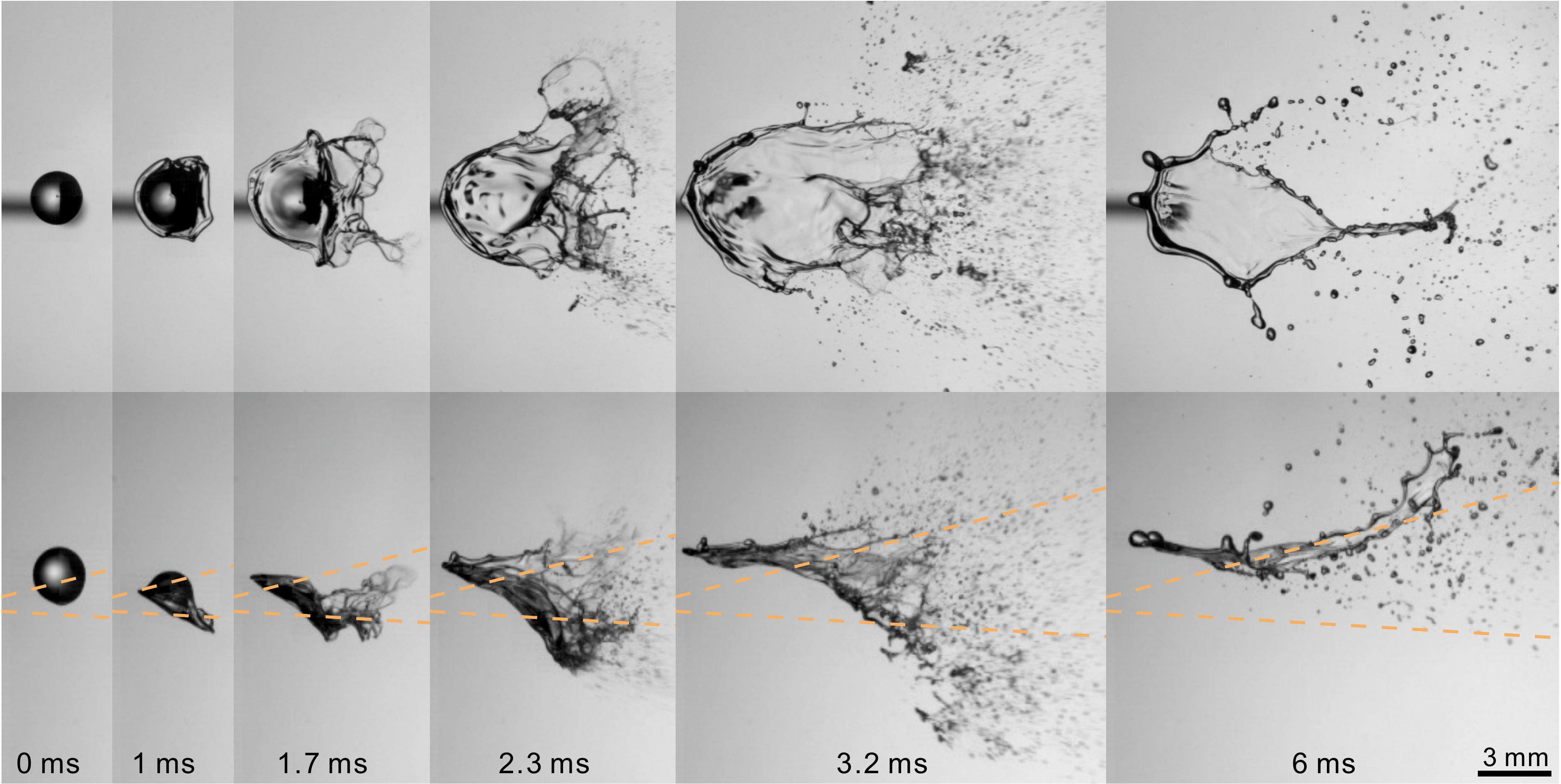}}
  \caption{Synchronised image sequences of droplet breakup from the bottom view (first row) and the side view (second row) with the conditions $\We_g = 230$, $\We_d =36.5$, $ \chi = 0.399$, ethanol droplet diameter $d_0 = 2.4$ mm. The orange lines indicate the velocity boundaries of the shear layer, corresponding to 5--95\% of the jet velocity. The breakup regime is the swing breakup with rim shrinkage. The corresponding movie can be found in the supplementary material (movie 7).}
\label{fig:fig17}
\end{figure}

\begin{figure}
  \centerline{\includegraphics[width=1\columnwidth]{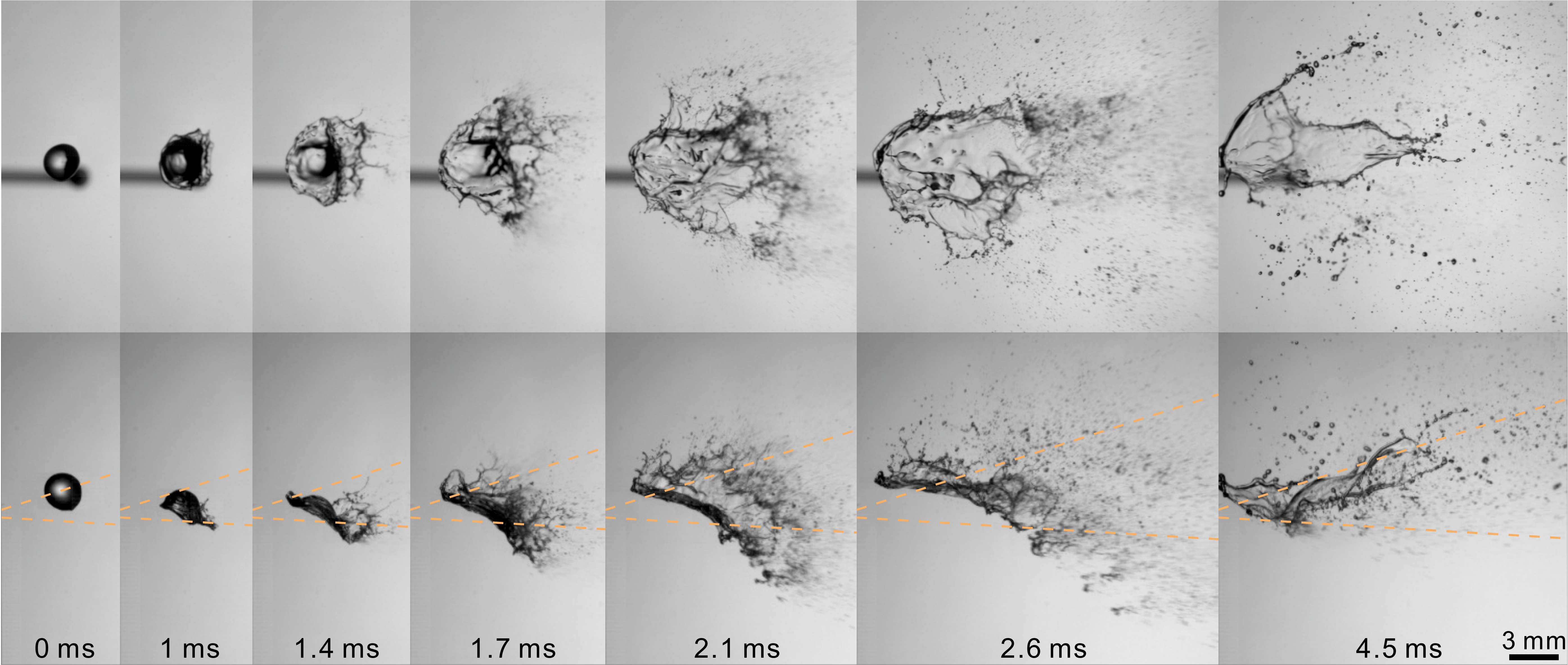}}
  \caption{Synchronised image sequences of droplet breakup from the bottom view (first row) and the side view (second row) with the conditions $\We_g =408$, $\We_d = 65.4$, $ \chi = 0.400$, ethanol droplet diameter $d_0 = 2.4$ mm. The orange lines indicate the velocity boundaries of the shear layer, corresponding to 5--95\% of the jet velocity. The breakup regime is the swing breakup with rim collapse. The corresponding movie can be found in the supplementary material (movie 8).}
\label{fig:fig18}
\end{figure}

\begin{figure}
  \centerline{\includegraphics[width=0.8\columnwidth]{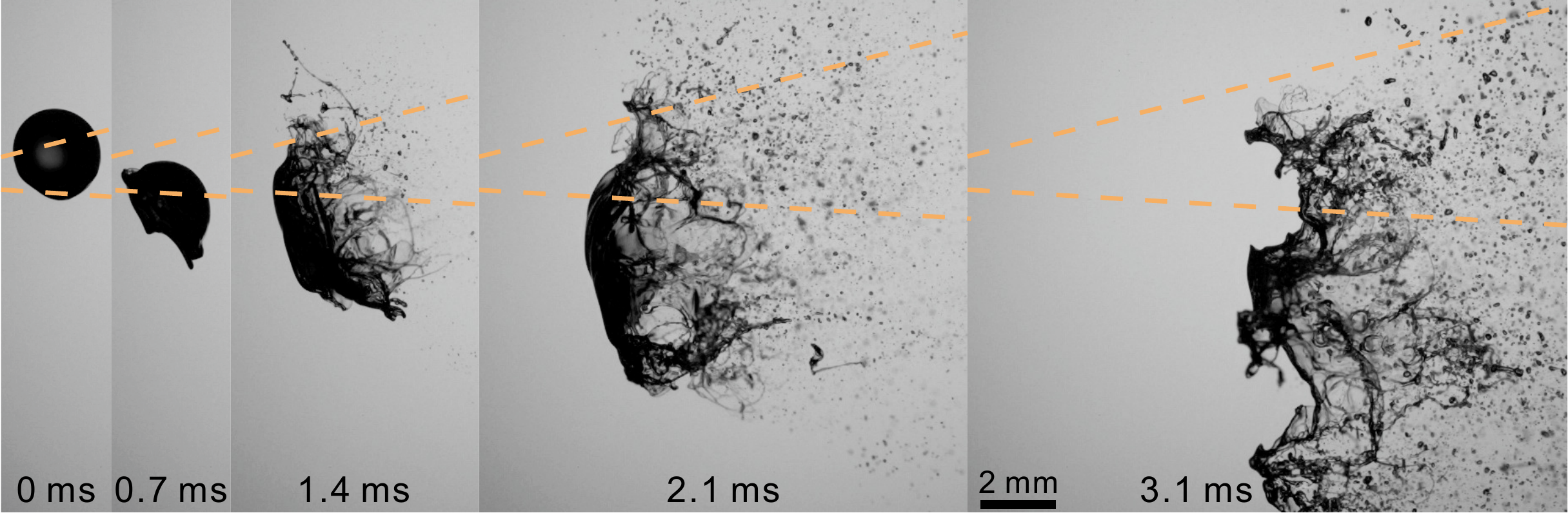}}
  \caption{Side-view images with the conditions $\We_g = 230$, $\We_d = 312$, $\chi = 1.165$, ethanol droplet diameter $d_0 = 2.4$ mm. The orange lines indicate the velocity boundaries of the shear layer, corresponding to 5--95\% of the jet velocity. The breakup regime is the shear-stripping breakup. The corresponding movie can be found in the supplementary material (movie 9).}
\label{fig:fig19}
\end{figure}

Three image sequences of droplet breakup under different parameters are provided in figures \ref{fig:fig17}--\ref{fig:fig19}. Figure \ref{fig:fig17} is for the swing breakup with rim shrinkage, figure \ref{fig:fig18} is for the swing breakup with rim collapse, and figure \ref{fig:fig19} is for the shear-stripping breakup.

\section{Comparison with the flapping breakup mechanism}\label{sec:AppB}
The swing breakup of the droplet body is similar to the flapping instability observed in the breakup of a liquid jet with a coaxial airflow. In both cases, the liquid deformation causes the liquid surface to be perpendicular to the flow direction of airflow, which further induces transverse destabilisation. However, the causes of the liquid deformation in the two cases are different.

In a coaxial liquid jet, the liquid surface is initially parallel to the airflow, and the initial liquid deformation originates from the shear instability \citep{Delon2018Flapping, Villermaux2004Transverse}. In this regard, we attempt to use the shear instability to analyse the droplet deformation. We can use the momentum thickness of the jet shear layer ($\delta _2$) as the gas boundary layer thickness at the droplet interface, and $\delta _2 \approx 0.5$ mm at $x = 5$ mm, which is the position of the windward surface of the droplet. We can obtain that $\We _\delta \approx $ 21--191, where ${{\We}_{\delta }}={{{\rho }_{g}}u_{g}^{2}{{\delta }_{2}}}/{\sigma }$ is the Weber number based on the gas boundary layer thickness. Therefore, the current case corresponds to the Rayleigh limit of the shear instability ($\We_{\delta }\gg 1$) \citep{Eggers2008ShearatInterface}. According to the analysis of shear instability at the Rayleigh limit \citep{Villermaux2004Transverse}, the most-amplified wavelength is
\begin{equation}\label{eq:eq27}
  {{\lambda }_{\KH}}\simeq \frac{2\pi }{1.5}{{\left( \frac{{{\rho }_{d}}}{{{\rho }_{g}}} \right)}^{1/2}}{{\delta }_{2}}.
\end{equation}
In our case, ${{\lambda }_{\KH}}\approx 50$ mm, which is much larger than the droplet size. Therefore, for the swing breakup in our experiment, the shear instability based on the thickness of the shear layer cannot develop in the droplet length scale.

The shear instability waves, however, do appear on the windward surface of the droplet, especially for high $\We_g > 230$. Considering the shear instability in the droplet length scale, we can get the most-amplified wavelength based on the KH limit of shear instability: \citep{Villermaux2004Transverse}
\begin{equation}\label{eq:eq28}
  {{\lambda }_{\KH,1}}=\frac{3\pi \sigma }{{{\rho }_{g}}u_{g}^{2}}.
\end{equation}
From (\ref{eq:eq28}), we can obtain ${{\lambda }_{\KH,1}}<0.1$ mm in our experiment, indicating the possibility of the shear instability wave in the droplet length scale. Although the predicted wavelength of shear instability based on the KH limit underestimates the actual wavelength in many studies \citep{Jalaal2014DropletInstabilities, Sharma2021Aerobreakup}, we can still assume ${{\lambda }_{\KH,1}}\ll {{d}_{0}}$ based on the experimental observations, as shown in figure \ref{fig:fig06}a. The small-scale shear instability waves tend to cause local peeling on the droplet surface, or merge into a large liquid tongue. Hence, they cannot induce an intensive swing of the droplet body.

\section{Mean and variance of size distributions}\label{sec:AppC}
From (\ref{eq:eq17}), the mean and variance of the log-normal distribution are
\begin{equation}\label{eq:eq29}
  {{\EX}_{\log }}={{e}^{\ln {{d}_{m}}-{\sigma _{\log }^{2}}/{2}\;}},\quad
  {{\DX}_{\log }}=\left( {{e}^{\sigma _{\log }^{2}}}-1 \right){{e}^{2\ln {{d}_{m}}+\sigma _{\log }^{2}}}.
\end{equation}
From (\ref{eq:eq18}), the mean and variance of the infinitely divisible distribution are
\begin{equation}\label{eq:eq30}
 {{\EX}_{\inf }}={{d}_{\inf }},\quad {{\DX}_{\inf }}=\sigma _{\inf }^{2}.
\end{equation}
From (\ref{eq:eq19}), the mean and variance of the gamma distribution are
\begin{equation}\label{eq:eq31}
 {{\EX}_{g}}=\alpha \beta ,\quad
 {{\DX}_{g}}=\alpha {{\beta }^{2}}.
\end{equation}
The mean and variance of the compound distribution can be obtained from the elementary gamma distributions
\begin{equation}\label{eq:eq32}
{{\EX}_{c}}=\sum\limits_{i=1}^{2}{{{A}_{i}}{{\EX}_{g,i}}},\quad
{{\DX}_{c}}=\sum\limits_{i=1}^{2}{{{A}_{i}}\left( \EX_{g,i}^{2}+{{\DX}_{g,i}} \right)}-\EX_{c}^{2},
\end{equation}
where $\EX_{g,i}$ and $\DX_{g,i}$ are the means and variances of the elementary gamma distributions.

The mean and variance of the volume-weighted size distribution of the experimental results are \citep{Lefebvre2017Atomization}
\begin{equation}\label{eq:eq33}
 {{\EX}_{ex}}={{d}_{43}},\quad
 {{\DX}_{ex}}=d_{53}^{2}-d_{43}^{2},
\end{equation}
where $d_{43}$ and $d_{53}$ are the representative diameters, calculated by
\begin{equation}\label{eq:eq34}
 {{d}_{pq}}={{\left( \frac{\sum{d_{i}^{p}}}{\sum{d_{i}^{q}}} \right)}^{1/\left( p-q \right)}},
\end{equation}
where $d_i$ is the size of each fragment, and $p$ and $q$ take on the values corresponding to $d_{43}$ and $d_{53}$.

\bibliographystyle{jfm}
\bibliography{dropletBreakup-all}
\end{document}